\documentclass[11pt,x11names]{article}

\usepackage[T1]{fontenc}
\usepackage[latin1]{inputenc}

\usepackage{fancyhdr,cancel}

\usepackage{listings}
\usepackage{enumitem,multicol,xcolor}

\usepackage{libertine}

\usepackage[hang,small,labelfont=bf,up,textfont=it,up]{caption}

\usepackage{amsmath,amsfonts,amsthm,amssymb,bm,dsfont}	

\DeclareMathOperator{\rank}{rank}

\usepackage{setspace}

\usepackage{graphicx}
\usepackage{subcaption}

\usepackage{adjustbox}

\usepackage[a4paper,top=3cm,bottom=3cm,left=2.3cm,right=2.3cm]{geometry}

\usepackage{dutchcal}
\usepackage{centernot}

\usepackage{numprint}
\npthousandsep{,}\npthousandthpartsep{}\npdecimalsign{.}

\renewcommand{\vec}[1]{\boldsymbol{#1}}
\newcommand{\mvec}[1]{\mathbf{#1}}
\newcommand{\abs}[1]{\vert{#1}\vert}

\renewcommand{\bf}{\bfseries}
\newcommand{\dplus}{d_+}
\newcommand{\dminus}{d_-}
\newcommand{\dtplus}{\check{d}_+}
\newcommand{\dtminus}{\check{d}_-}

\newcommand{\Xd}{\mvec X_{:d}}
\newcommand{\Xdd}{\mvec X_{d:}}
\newcommand{\Xk}{\mvec X^{(k)}}
\newcommand{\Xkd}{\Xk_{:d}}
\newcommand{\Xkdd}{\Xk_{d:}}
\newcommand{\xid}{\vec x_{i,:d}}
\newcommand{\xidd}{\vec x_{i,d:}}
\newcommand{\xjd}{\vec x_{j,:d}}
\newcommand{\xjdd}{\vec x_{j,d:}}
\newcommand{\m}{\vec m^{(k)}_{:d}}

\newcommand{\D}{\mvec D^{(k)}_{:d}}
\newcommand{\Dki}{\mvec D^{(k),-i}_{:d}}
\newcommand{\Dzero}{\bm\Delta_d}
\newcommand{\s}{s^{(k)}_j}
\newcommand{\sdotkj}{s^{(\bullet k)}_j}
\newcommand{\ski}{s^{(k),-i}_j}

\newcommand{\Xs}{\mvec X}
\newcommand{\Xr}{\mvec X^\prime}

\newcommand{\titledoc}{Bayesian estimation of the latent dimension and communities in stochastic blockmodels}
\newcommand{\titleshort}{Bayesian estimation of the latent dimension and number of communities}

\usepackage{csquotes} 
\usepackage{natbib}

\usepackage{sectsty}
\partfont{\fontsize{17}{17}\selectfont}

\usepackage{thmtools}
\newtheorem{definition}{Definition}

\usepackage{authblk}

\usepackage[ruled,linesnumbered]{algorithm2e}
\SetKwInput{KwInput}{Input}
\usepackage{hyperref}
\hypersetup{colorlinks,linkcolor={blue},citecolor={blue},urlcolor={blue}}  
\numberwithin{equation}{section}

\usepackage{mathtools}
\mathtoolsset{showonlyrefs}

\author{Francesco Sanna Passino and Nicholas A. Heard \\ 
Department of Mathematics, Imperial College London \\ 180 Queen's Gate, SW7 2AZ, London, United Kingdom
}

\date{\textit{This is a pre-print of an article published in Statistics and Computing. The final authenticated version is available online at: \url{https://doi.org/10.1007/s11222-020-09946-6}.}}

\title{\huge\textbf{\titledoc}}
	
\begin{document}


\maketitle



\begin{abstract}
Spectral embedding of adjacency or Laplacian matrices of undirected graphs is a common technique for representing a network in a lower dimensional latent space, with optimal theoretical guarantees. 
The embedding can be used to estimate the community structure of the network, with strong consistency results in the stochastic blockmodel framework. 
One of the main practical limitations of standard algorithms for community detection from spectral embeddings is that the number of communities and the latent dimension of the embedding must be specified in advance. 
In this article, a novel Bayesian model for simultaneous and automatic selection of the appropriate dimension of the latent space and the number of blocks is proposed. 
Extensions to directed and bipartite graphs are discussed. 
The model is tested on simulated and real world network data, showing promising performance for recovering latent community structure.
\end{abstract}

\section{Introduction}

A network can be represented as a graph $\mathbb G=(V,E)$, where $V$ is a set of nodes and $E\subseteq V\times V$ is a set of edges indicating the pairs of nodes which have interacted. 
The graph can be characterised by the adjacency matrix $\mvec A\in\{0,1\}^{n\times n}$, where $n=\abs{V}$ and for $1\leq i,j\leq n$, $A_{ij}=\mathds 1_E\{(i,j)\}$, such that $A_{ij}=1$ if a link between the nodes $i$ and $j$ exists, and $A_{ij}=0$ otherwise. The graph is said to be undirected if $(i,j)\in E \iff (j,i) \in E$ and $\mvec A$ is constrained to be symmetric; otherwise, the graph is said to be directed. It will be assumed that a node cannot link to itself, implying $\mvec A$ is a hollow matrix.

Latent space models \citep{hoff,handcock,krivitsky} represent a flexible approach to statistical analysis of networks: each node $i$ is assigned a latent position $\vec x_i$ in a $d$-dimensional latent space $\mathbb X$, and edges between pairs of nodes are typically generated independently (\textit{independent-edge} graph), with the probability of observing a link between nodes $i$ and $j$ obtained through a \textit{kernel} function $\psi:\mathbb X\times\mathbb X\to[0,1]$ of the respective latent positions: 
$\mathbb P(A_{ij}=1)=\psi(\vec x_i,\vec x_j)$. 
Different ideas and techniques for embedding observed graphs into low dimensional spaces are explored in the literature \citep[for a survey, see, for example][]{cai_survey}. 
\textit{Random dot product graphs} (RDPGs) \citep{Nickel06,young_rdpg,young} are a popular class of latent position models, 
where $\mathbb X\subseteq\mathbb R^d$, and the function $\psi(\cdot)$ is an inner product $\langle\cdot,\cdot\rangle$ on $\mathbb X\times\mathbb X$. 
RDPGs are analytically tractable and have therefore been extensively studied; a survey of the existing statistical inference techniques is presented in \citet{athreya}. 

The \textit{stochastic blockmodel} (SBM) \citep{holland} is the classical statistical model for clustering graphs \citep{snijders,nowi}. 
Assuming $K$ communities, each node is assigned a community membership $z_i\in\{1,\dots,K\}$ with probabilities $\bm\theta$, from the $K-1$ probability simplex.
The probability of a link only depends on the community allocations $z_i$ and $z_j$ of the two nodes. 
Given a symmetric matrix $\mvec B\in[0,1]^{K\times K}$ of inter-community probabilities, then independently $\mathbb P(A_{ij}=1)=B_{z_iz_j}$. The likelihood for an observed symmetric adjacency matrix $\mvec A$ is therefore
\begin{equation}
L(\mvec A\vert\vec z,\mvec B)=\prod_{1\leq i<j\leq n} B_{z_iz_j}^{A_{ij}}(1-B_{z_iz_j})^{1-A_{ij}}. \label{eq:lhd}
\end{equation}
Stochastic blockmodels have appealing statistical properties, and can well approximate any independent-edge network model if the number of communities is sufficiently large \citep{Bickel09,olhede}. 
Limits and bounds for community detection in stochastic blockmodels, from an information-theoretic approach, have been extensively studied and explored in the literature \citep{abbe_recovery,abbe}. 
SBMs have also been successfully extended in different directions: 
allowing for multiple overlapping communities \citep{airoldi}, 
incorporating degree corrections \citep{karrer}, 
capturing node popularity \citep{sengupta}, 
Bayesian models \citep{peixoto,vdv} 
and dynamically evolving models \citep{xu,matias,pensky,ludkin}. 
Stochastic blockmodels can also be easily represented as random dot product graphs: 
each community is assigned a latent position, which is common to all the nodes belonging to the cluster, and $\mvec B$ is obtained from the inner products of those positions. Hence, in this framework, $d=\rank(\mvec B)\leq K$.

Spectral clustering \citep{spectral} provides a consistent statistical estimation procedure for the latent positions in SBMs \citep{rohe,sussman,Fishkind13,lyzinski,lei,lyz_community} 
and more generally in random dot product graphs \citep{Tang13,sussman_ieee}. 
Community memberships in the stochastic block model can be consistently estimated using $K$-means clustering of the rows \citep{rohe}. 
\citet{prd} directly links spectral embedding to the generalised random dot product graph (GRDPG), an extension of the RDPG, 
and advocates for Gaussian mixture modelling (GMM) of the rows of the embedding, especially when the Laplacian matrix is used. 
Alternatives to spectral clustering include variational methods \citep{celisse} and pseudo-likelihood approaches \citep{amini}. 
SBMs have been extended to the directed case \citep{wang_wong,scbm}, and appropriate embeddings for co-clustering, 
in most cases based on the singular value decomposition (SVD), have been derived in the literature \citep{dhillon,malliaros,zheng}. 

One of the practical issues of spectral embedding, and in general all graph embedding algorithms \citep{cai_survey}, 
is that it requires a suitable prespecified latent dimensionality $d$ (usually $d\ll\abs{V}$) as input, 
and, subsequently, a suitable number of clusters $K$, conditionally, crucially, on the previous choice of $d$. 
For a practical example of this procedure on a real world network, see \cite{Priebe19}. 
In general, in spectral clustering, similarly to what practitioners do in principal component analysis (PCA), 
the investigator examines the scree-plot of the eigenvalues and chooses the dimension based on the location of \textit{elbows} in the plot \citep{jolliffe},
or uses the \textit{eigengap} heuristic \citep[see, for example,][]{spectral}. 
Automatic methods for thresholding have also been suggested \citep{zhu,chatter}.
A relevant body of literature is also devoted to methods for the selection of the number of communities in stochastic blockmodels
\citep{Zhao11,Bickel16,Lei16,reinert,Saldana17,riolo,wang_model_selection,Chen18,rastelli}. 
Often, practitioners simply set $d=K$, for some $d$, assuming that $\mvec B$ has full rank in the stochastic blockmodel framework. 
Under the full rank assumption, one may estimate $d=K$ as the number of eigenvalues of $\mvec A$ which are larger than $\sqrt{n}$ \citep{chatter,Lei16}. 
In this setting, hypothesis tests have also been proposed \citep{Zhao11,Bickel16,Lei16}.
In this work, the problem of selecting $d$ is approached from the perspective of variable selection in model based clustering \citep{raftery,lau}, which is widely studied in the literature \citep{fowlkes,law,dean,maugis,andrews}. 

Similarly, the problem of correctly selecting the number of clusters is also common in $K$-means or GMMs, since it is usually required to specify a number of components in the mixture. 
Usually the parameter is chosen by minimising information criteria (for example, AIC or BIC). 
Numerous other techniques have been proposed for GMMs with unknown number of components \citep{mengersen,richardson,stephens,nobile,zhang,dellaportas,nobile_fearn,miller_mix}.

Clearly, the sequential approach in estimating $d$ and $K$ is suboptimal, and it is desirable to jointly estimate the two parameters. This article addresses the problem in a Bayesian framework, proposing a novel methodology to automatically select $d$ and $K$, simultaneously.
Techniques for selection of $K$ in GMMs will be incorporated within the spectral embedding framework, allowing for
$K$ and $d$, the number of communities and latent dimension of the latent positions, to be random and learned from the data.
A structured Bayesian model for simultaneously inferring the dimension of the latent space, the number of communities, and the community allocations is proposed. 
The model is based on asymptotic results \citep{tang,prd} on the leading and informative components of spectral embeddings, obtained for $d$ fixed and known. 
The asymptotic theory is combined with realistic assumptions about the remaining components of the embedding, empirically tested and justified on simulated data.
Furthermore, extensions to the directed and bipartite case will be discussed. 
The proposed model has multiple advantages: the latent dimension $d$ and number of communities $K$ are modelled separately, and the Bayesian framework allows for automatic selection of the two parameters. 
The model also allows estimation of $d$ even when $d<K$, and gives insights on the goodness-of-fit of the stochastic blockmodel on observed network data, based on the embedding. 
The method is tested on simulated data and applied to real world computer and transportation networks. It should be noted that \cite{Yang19} have simultaneously and independently proposed a similar inferential procedure within a frequentist framework.

The article is organised as follows: Section~\ref{grdpg_sec} introduces adjacency and Laplacian spectral embeddings and the GRDPG. 
The novel Bayesian model for selection of the appropriate dimension of the latent space is discussed in Section~\ref{sec:model_sec}, followed by careful illustration of posterior inference procedures in Section~\ref{mcmc_inference}. 
Section~\ref{curse_section} discusses the effects of the curse of dimensionality on the model, and suggests a remedy.
Extensions of the model are presented in Section~\ref{ext}, and results and applications are finally discussed in Section~\ref{results}.

\section{GRDPG and spectral embeddings} \label{grdpg_sec}

The adjacency matrix of an undirected graph 
is usually embedded into a latent space of dimension $d$ using one of two different procedures: the adjacency spectral embedding or the Laplacian spectral embedding. Suppose $\mvec A\in\{0,1\}^{n\times n}$ is a symmetric adjacency matrix of an undirected graph with $n$ nodes.

\begin{definition}[Adjacency spectral embedding -- ASE] \label{adj_emb}
For $d\in\{1,\ldots,n\}$, consider the spectral decomposition
\begin{equation}
\mvec A = \hat{\bm\Gamma}\hat{\bm\Lambda}\hat{\bm\Gamma}^\top +  \hat{\bm\Gamma}_\perp\hat{\bm\Lambda}_\perp\hat{\bm\Gamma}_\perp^\top,
\end{equation}
where $\hat{\bm\Lambda}$ is a $d\times d$ diagonal matrix containing the top $d$ eigenvalues in magnitude, in decreasing order, $\hat{\bm\Gamma}$ is a $n\times d$ matrix containing the corresponding orthonormal eigenvectors, and the matrices $\hat{\bm\Lambda}_\perp$ and $\hat{\bm\Gamma}_\perp$ contain the remaining $n-d$ eigenvalues and eigenvectors. The adjacency spectral embedding $\hat{\mvec X}=[\hat{\vec x}_{1},\dots,\hat{\vec x}_{n}]^\top$ of $\mvec A$ in $\mathbb R^d$ is 
\begin{equation}
\hat{\mvec X} = \hat{\bm\Gamma}\vert\hat{\bm\Lambda}\vert^{1/2}\in\mathbb R^{n\times d},
\end{equation}
where the operator $\vert\cdot\vert$ applied to a matrix returns the absolute value of its entries.
\end{definition}

\begin{definition}[Laplacian spectral embedding -- LSE] \label{lap_emb}
  For $d\in\{1,\ldots,n\}$, consider the (modified) normalised Laplacian matrix
\begin{equation}
\mvec L = \mvec D^{-1/2}\mvec A\mvec D^{-1/2},\ 
\mvec D = \mathrm{diag}\left(\sum\nolimits_{j=1}^n A_{ij}\right),
\end{equation}
and its spectral decomposition
\begin{equation}
\mvec L = \tilde{\bm\Gamma}\tilde{\bm\Lambda}\tilde{\bm\Gamma}^\top +  \tilde{\bm\Gamma}_\perp\tilde{\bm\Lambda}_\perp\tilde{\bm\Gamma}_\perp^\top.
\end{equation}
The Laplacian spectral embedding $\tilde{\mvec X}=[\tilde{\vec x}_{1},\dots,\tilde{\vec x}_{n}]^\top$ of $\mvec A$ in $\mathbb R^d$ is 
\begin{equation}
\tilde{\mvec X} = \tilde{\bm\Gamma}\vert\tilde{\bm\Lambda}\vert^{1/2}.
\end{equation}
\end{definition}

The modified Laplacian $\mvec D^{-1/2}\mvec A\mvec D^{-1/2}$ \citep{rohe} is preferred to the more common $\mvec I_n - \mvec D^{-1/2}\mvec A\mvec D^{-1/2}$ since the eigenvalues of the former lie in $(-1,1)$, providing a convenient interpretation for disassortative networks \citep{disassortativity}. Also, using the Laplacian embedding is preferable in sparse regimes, whereas the adjacency embedding should be used for relatively dense graphs \citep{Cape18}. A case-study with discussion on the differences between community structures obtained from ASE and LSE is presented in \cite{Priebe19}.

In this work, the stochastic block model will be interpreted as a specific case of a generalised random dot product graph (GRDPG) \citep{prd}. The GRDPG, a generic latent position model for graphs, is defined below.

\begin{definition}[Generalised random dot product graph, GRDPG] \label{grdpg}
Let $\dplus,\dminus$ be non-negative integers such that $d=\dplus+\dminus$. Let $\mathbb X\subseteq\mathbb R^{d}$ such that $\forall\ \vec x,\vec x^\prime\in\mathbb X$, $0\leq \vec x^\top{\mvec I}(\dplus,\dminus)\vec x^\prime\leq 1$, where
\begin{equation}
\mvec I(p,q) = \mathrm{diag}(\underbrace{1,\ldots,1}_{\displaystyle{p}},\underbrace{{-1},\ldots,{-1}}_{\displaystyle{q}}).
\end{equation}
Let $\mathcal{F}$ be a probability measure on $\mathbb{X}$, $\mvec A\in\{0,1\}^{n\times n}$ be a symmetric matrix and $\mvec{X}=(\vec x_1,\dots,\vec x_n)^\top\in\mathbb{X}^n$. Then $(\mvec A,\mvec X)\sim\mathrm{GRDPG}_{\dplus,\dminus}(\mathcal{F})$ if $\vec x_1,\dots,\vec x_n\overset{iid}{\sim}\mathcal F$ and for $i<j$, independently
\begin{equation}
\mathbb P(A_{ij}=1)=\vec x_i^\top{\mvec I}(\dplus,\dminus)\vec x_j.
\end{equation}
\end{definition}

To represent the $K$-community stochastic blockmodel as a GRDPG model, $\mathcal{F}$ can be chosen to have mass concentrated at $\bm \mu_1,\dots, \bm \mu_K \in\mathbb R^{d}$ such that $\vec \mu_i^\top{\mvec I}(\dplus,\dminus)\vec \mu_j=B_{ij}\ \forall\ i,j\in\{1,\dots,K\}$. 
For estimation of the latent positions in a SBM, interpreted as a GRDPG, \citet{prd} suggest the following algorithm, based on Gaussian mixture modelling \citep[see, for example,][]{raftery}.

\begin{algorithm}[!h]
\SetAlgoLined
\KwInput{adjacency matrix $\mvec A$ (or Laplacian $\mvec L$), dimension $d$, number of communities $K\geq d$.}
compute spectral embeddings $\hat{\mvec X}=[\hat{\vec x}_{1},\dots,\hat{\vec x}_{n}]^\top$  (or $\tilde{\mvec X}=[\tilde{\vec x}_{1},\dots,\tilde{\vec x}_{n}]^\top$) into $\mathbb R^d$, \\
fit a Gaussian mixture model to $\hat{\mvec X}$ (or $\tilde{\mvec X}$) with $K$ components. \\
\KwResult{return cluster centres $\vec \mu_1,\dots,\vec \mu_K\in\mathbb R^d$ and cluster allocations $z_1,\dots,z_n$.}
 \caption{Spectral estimation of the SBM (spectral clustering)}
 \label{sbm_est}
\end{algorithm}

Intuitively, the algorithm approximately holds because, taking a graph with $m$ nodes, and restricting the attention to the first $n$ nodes, with $n<m$:
\begin{equation}
\mvec Q_m\hat{\vec x}_{i} \overset{d}{\longrightarrow} \mathbb N\{\vec \mu_{z_i}, m^{-1/2}\bm\Sigma(\vec \mu_{z_i})\},\ m\to\infty,\ i=1,\dots,n,
\end{equation}
where $\mvec Q_m$ is a matrix from the indefinite orthogonal group $\mathbb O(\dplus,\dminus)$ and $\bm\Sigma(\vec \mu_{z_i})$ can be analytically computed \citep{prd}. 
The result holds for $d$ fixed and known, but in this work it is of interest to treat $d$ as a random, unknown parameter. 
If a $m$-dimensional embedding is considered, with $m>d$, then asymptotic theory implies an approximate normal distribution with non-zero means and an elliptic covariance within each cluster for the top-$d$ components of the embedding; but, \textit{to the best of our knowledge}, no theoretical results have been obtained for the remaining $m-d$ columns.
It is therefore necessary to propose a model for the remaining part of the embedding, which will be carefully described in Section~\ref{sec:model_sec}, and empirically  justified and assessed in Section~\ref{section_vali}.

It should be noted that it is only possible to estimate the vectors $\{\vec \mu_j\}$ up to an orthogonal transformation; specifically, for any matrix $\mvec Q\in\mathbb O(\dplus,\dminus)$, 
$(\mvec Q\vec \mu_{z_i})^\top{\mvec I}(\dplus,\dminus)(\mvec Q\vec \mu_{z_j})=\vec \mu_{z_i}^\top{\mvec I}(\dplus,\dminus)\vec \mu_{z_j}$, which implies that the likelihood \eqref{eq:lhd} is invariant to any such transformation. This is inconsequential for much the same reason, however, as knowledge of the true transformation would not change any inferences about the network.


A second source of non-identifiability in the GRDPG interpretation of the SBM is the \textit{"uniqueness up to artificial dimension blow-up"} \citep{Cape18}: for $(\mvec A,\mvec X)\sim\mathrm{GRDPG}_{\dplus,\dminus}(\mathcal{F})$, there exists $\check{\mathcal F}$ on $\mathbb R^{\check{d}}$, with $\check{d} > d$, such that $(\mvec A,\mvec X)\overset{d}{=}(\check{\mvec A},\check{\mvec X})$ with $(\check{\mvec A},\check{\mvec X})\sim\mathrm{GRDPG}_{\dtplus,\dtminus}(\check{\mathcal{F}})$. In the stochastic blockmodel setting, this essentially means that any matrix $\mvec B\in[0,1]^{K\times K}$ with rank $d$ can be obtained as an inner product between latent positions on arbitrarily large dimensions.


\section{A Bayesian model for SBM embeddings} \label{sec:model_sec}

For simplicity, the embeddings will be generically denoted as ${\mvec X}=[\vec x_1,\dots,\vec x_n]^\top\in\mathbb R^{n\times m},\ \vec x_i\in\mathbb R^m$ for some $m$, $d\leq m\leq n$. In this article, $m$ is always assumed to be fixed and obtained from a preprocessing step. Choosing an appropriate value of $m$ is arguably much easier than choosing the correct $d$, and, in the proposed model, the correct $d$ can be recovered for any choice of $m$, as long as $d\leq m$. Let $\Xd$ denote the first $d$ columns of ${\mvec X}$, and $\Xdd$ the $m-d$ remaining columns. 
The notation $\xid$ denotes the first $d$ elements $(x_1,\dots,x_d)$ of the vector $\vec x_i$, and similarly $\xidd$ denotes the last $m-d$ elements $(x_{d+1},\dots,x_{m})$.

Suppose a latent dimension $d$, $K$ communities, and latent community assignments $\vec z=(z_1,\dots,z_n)$. The latent positions of nodes within each community are assumed to be generated from an $m$-dimensional community-specific Gaussian distribution, where the first $d$ components are modelled differently from the remaining $m-d$: 
the initial components $\xid$ are assumed to have unconstrained mean vector $\bm\mu_k\in\mathbb{R}^d$ and positive definite $d\times d$ covariance matrix $\bm\Sigma_k$; in contrast, for $\xidd$, two constraints are imposed: the mean is an $(m-d)$-dimensional vector of zeros, 
and the covariance is a diagonal matrix $\bm\sigma^2_k\mvec I_{m-d}$ with positive entries $\bm\sigma^2_k=(\sigma^2_{k,d+1},\dots,\sigma^2_{k,m})$. The validation of the model assumptions will be discussed in Section~\ref{section_vali}. For mathematical convenience, conjugate priors can be placed on the parameters as follows:
\begin{align}
\vec x_i \vert d,z_i,\vec \mu_{z_i},\bm\Sigma_{z_i}, \bm\sigma^2_{z_i} &\overset{d}{\sim} \mathbb N_m \left( \begin{bmatrix} \vec \mu_{z_i} \\ \vec 0 \end{bmatrix}, \begin{bmatrix} \bm\Sigma_{z_i} & \vec 0 \\ \vec 0 & \bm\sigma^2_{z_i}\mvec I_{m-d} \end{bmatrix} \right),\ i=1,\dots,n, \\
(\vec \mu_{k},\bm\Sigma_{k})\vert d &\overset{iid}{\sim} \mathrm{NIW}_d(\vec 0, \kappa_0,\nu_0+d-1,\Dzero),\ k=1,\dots,K, \\
\sigma^{2}_{k,j} &\overset{iid}{\sim} \mathrm{Inv}\text{-}\chi^2(\lambda_0,\sigma_{0}^2),\ j=d+1,\dots,m, \\
d\vert\vec z &\overset{d}{\sim} \mathrm{Uniform}\{1,\dots,K_\varnothing\}, \\
z_i\vert\bm\theta &\overset{iid}{\sim}\mathrm{Multinoulli}(\bm\theta),\ i=1,\dots,n, 
\\
\bm\theta \vert K &\overset{d}{\sim} \mathrm{Dirichlet}\left({\alpha}/{K},\dots,{\alpha}/{K}\right), \\
K &\overset{d}{\sim} \mathrm{Geometric}(\omega),
\label{full_model}
\end{align}
where if, $n_k= \sum\nolimits_{i=1}^n \mathds 1_k\{z_i\}$ is the size of community $k$, $K_\varnothing=\sum_{k=1}^K\mathds 1_{\mathbb N_+}\{n_k\}$ is the number of non-empty communities. 
Note that the inverse Wishart has been partially re-parametrised using a parameter $\nu_0>0$. 
The dimension of the corresponding matrix is then added to obtain the required constraint $\nu_0+d-1>d-1$ for the generic parametrisation and interpretation of the degrees of freedom $\nu$ (in this case $\nu=\nu_0+d-1$) of the distribution. 
Also, note that $m$ can be generically chosen to be equal to $K$, when fixed, or equal to $n$ to have the maximum possible dimension of the embedding. 
Since one of the specific objectives of the analysis is to learn the number of components of the mixture, rather than density estimation, the widely used infinite Bayesian nonparametric mixtures are not appropriate in this context \citep{harrison}. 

\cite{Yang19} have proposed a similar model in a frequentist framework. The conjecture on the distribution of $\mvec X_{d:}$ is essentially the same, except for the choice of the diagonal elements of the cluster-specific covariance matrix: \cite{Yang19} use a common variance parameter $\sigma^2_k$ for the last $m-d$ columns of the embedding, whereas a $(m-d)$-dimensional vector of variances $\bm\sigma^2_k$ is used in this paper. Additionally, as a second difference from \cite{Yang19}, the full model proposed here will also incorporate a second-level community cluster structure on these vectors of variances, which will be introduced in Section~\ref{curse_section}.

\begin{figure}[!t]
\centering
\includegraphics[width=.7\textwidth]{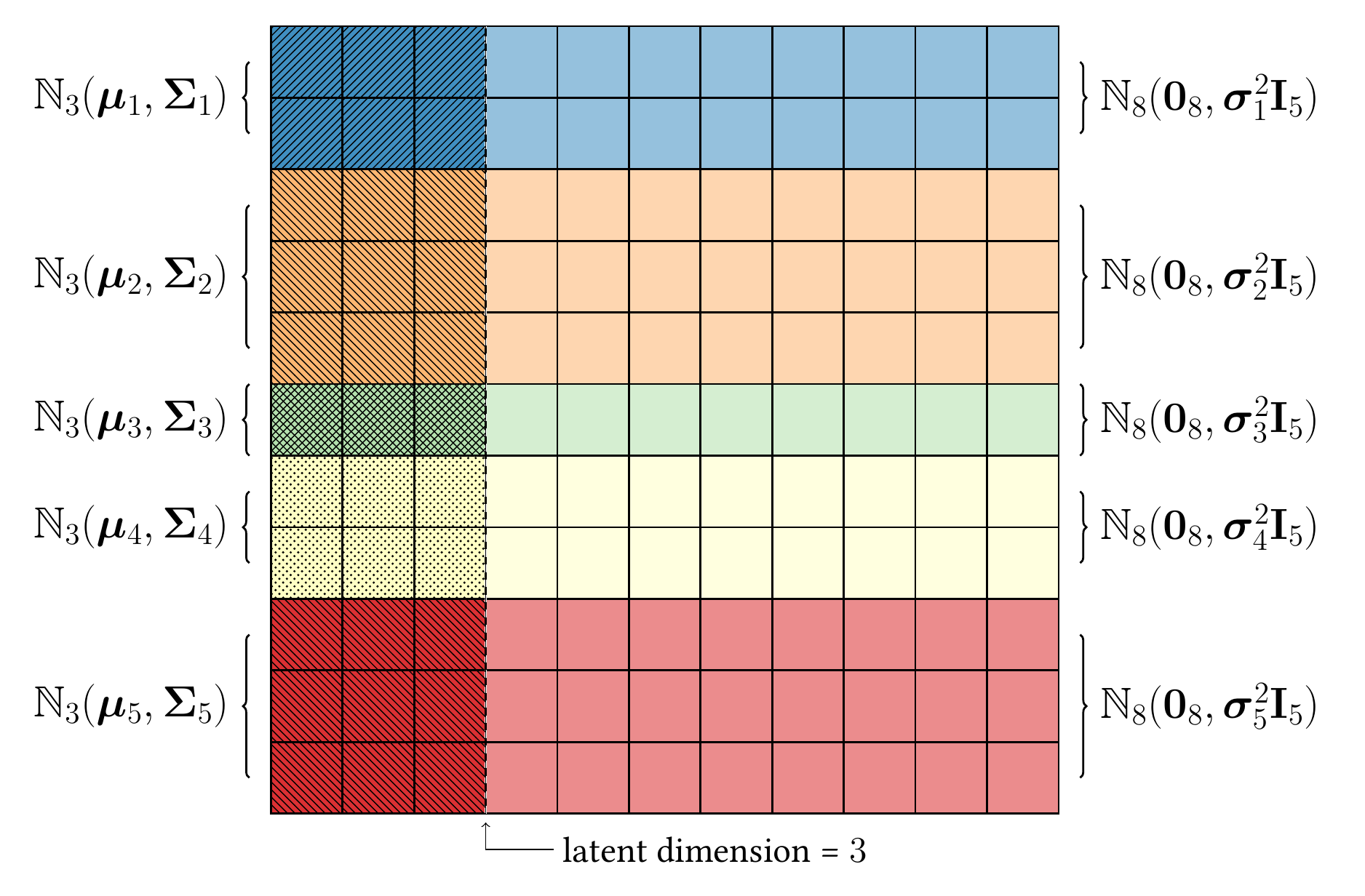}
\caption{Cartoon example for the generating process of the embedding of an 11-node stochastic blockmodel GRDPG with $K=5$ communities and latent dimension $d=3$.}
\label{fig:grid}
\end{figure}

A cartoon representation of the model is given in Figure~\ref{fig:grid}. Note that the condition $d\leq K$ is explicitly enforced in \eqref{full_model}. More specifically, $d\leq K_\varnothing$, which avoids an artificial matching between $d$ and $K$ using empty clusters, which are given non-zero probability mass under the Dirichlet-Multinoulli prior on $(\bm\theta,\vec z)$. 
One can also model $d$ and $K$ separately in an analogous way, changing the prior $p(d)$ to, for example,
\begin{equation}
  d \overset{d}{\sim}\mathrm{Geometric}(\delta),\label{eq:d_geom}
\end{equation}
  independently of $K$ and $\vec z$; this will later be referred to as the unconstrained model. 
The alternative prior \eqref{eq:d_geom} is particularly useful in practical applications and provides a useful interpretation of $d$: when $d\leq K$, then $d=\mathrm{rank}(\mvec B)$, but when $d>K$, this implies that the observed data might deviate from the stochastic blockmodel assumption, and provides a useful diagnostic for model validation and goodness-of-fit. 


The likelihood associated with the spectral embedding $\mvec X\in\mathbb R^{n\times m}$ obtained from a stochastic blockmodel can be expressed as:
\begin{equation}
  L({\mvec X}) =
\prod_{i=1}^n \left\{\phi(\xid;\vec \mu_{z_i},\bm\Sigma_{z_i})\prod_{j=d+1}^m \phi(x_{i,j};0,\sigma_{z_i,j}^2)\right\},
\end{equation}
where $\phi(\cdot)$ denotes the (possibly multivariate) Gaussian density function. 
Hence, the posterior, up to a normalising constant, has form
\begin{equation}
p(\{\bm\mu_k\},\{\bm\Sigma_k\},\{\bm\sigma^2_{k}\},\vec z,\bm\theta,K,d\vert \mvec X) \propto L(\mvec X) \prod_{k=1}^K \bigg\{p(\vec \mu_k,\bm\Sigma_k\vert d) \prod_{j=d+1}^{m} p(\sigma^2_{k,j}\vert d)\bigg\} \prod_{i=1}^n p(z_i\vert\vec\theta)p(K)p(d).
\end{equation} 

The $\mathrm{NIW}_d(\vec 0,\kappa_0,\nu_0+d-1,\bm\Delta_{d})$ prior on the pair $(\vec \mu_{k},\bm\Sigma_{k})$ is conjugate and yields a conditional posterior $(\vec \mu_{k},\bm\Sigma_{k})\vert\mvec X,\vec z,d\overset{d}{\sim}\mathrm{NIW}_d(\m,\kappa_{n_k},\nu_{n_k}+d-1,\D)$. By standard methods for inference in a multivariate Gaussian mixture model with $\mathrm{NIW}$ prior, the covariance matrix $\bm\Sigma_k$ can be explicitly integrated out from the posterior to obtain
\begin{equation}
p(\vec \mu_k\vert \mvec X,\vec z, d) = t_{\nu_{n_k}}\left(\bm\mu_k\vert\m,\D/(\kappa_{n_k}{\nu_{n_k}})\right), \label{margi_post}
\end{equation}
the density of the multivariate Student $t$ distribution with $\nu_{n_k}$ degrees of freedom, where
\begin{align}
\nu_{n_k} &=\nu_0+n_k, \\
\kappa_{n_k} &=\kappa_0+n_k, \\
\m &= \sum\nolimits_{i:z_i=k}\xid \big{/} \kappa_{n_k},\ \\
\D &= \Dzero + \sum\nolimits_{i:z_i=k} \xid\xid^\top - \kappa_{n_k}\m{\m}^\top. \label{post_mu}
\end{align}
Henceforth, $\vec \mu_k$ can easily be resampled in a simple Gibbs sampling step, conditional on the actual value of $d$ and on the community allocations $\vec z$. 
In this work, the location vectors $\vec \mu_k$ are collapsed out too, but the distribution is instructive to present other distributional results below, and could be also used if the objective of the analysis is to also recover the explicit form of the latent positions.

In a multivariate Gaussian model with normal inverse Wishart prior, it is also possible to analytically express the marginal likelihood of the observed data. Here, conditioning on a community-specific Gaussian, on the assignments $\vec z$ and on the dimension of the latent space $d$:
\begin{equation}
p(\Xkd\vert d,\vec z) = \pi^{-n_kd/2}\frac{\kappa_0^{d/2}\vert\Dzero\vert^{(\nu_0+d-1)/2}}{\kappa_{n_k}^{d/2}\vert\D\vert^{(\nu_{n_k}+d-1)/2}}\prod_{i=1}^d \frac{\Gamma\{(\nu_{n_k}+d-i)/2\}}{\Gamma\{(\nu_0+d-i)/2\}}, \label{marginal_left}
\end{equation}
where $\Xkd$ is the subset of rows of $\Xd$ such that $z_i=k$. 

 Given the $\mathrm{Inv}\text{-}\chi^2(\lambda_0,\sigma^2_0)$ prior, the posterior for $\sigma^2_{j,k}$ is $\mathrm{Inv}\text{-}\chi^2(\lambda_{n_k},\s)$, where
\begin{equation}
\lambda_{n_k} = \lambda_0 + n_k, \quad \s = \left\{\lambda_0\sigma^2_{0} + \sum\nolimits_{i:z_i=k} x_{ij}^2\right\}\Big /\lambda_{n_k}. \label{eq:sigma_post_pars}
\end{equation}
Similar calculations give the full marginal likelihood for the remaining portion of the embedding $\Xkdd$:
\begin{equation}
p(\Xkdd\vert d,\vec z) = \pi^{-n_k(m-d)/2} \bigg\{\frac{\Gamma(\lambda_{n_k}/2)}{\Gamma(\lambda_0/2)}\bigg\}^{m-d}\prod_{j=d+1}^{m}\frac{(\lambda_0\sigma^2_{0})^{\lambda_0/2}}{(\lambda_{n_k}\s)^{\lambda_{n_k}/2}}. \label{marginal_right}
\end{equation}
Finally, if $d$ is considered as a nuisance parameter:
\begin{equation}
p(\mvec X\vert\vec z,K) = \sum_{d=1}^K p(d\vert K)\prod_{k=1}^K p(\Xkd\vert d,\vec z)p(\Xkdd\vert d,\vec z),
\end{equation}
which is easily computed using \eqref{marginal_left} and \eqref{marginal_right}.

Also, note that the probabilities $\bm\theta$ associated to the community assignment can be easily integrated out, resulting in the following marginal likelihood, conditional on $K$:
\begin{equation}
p(\vec z\vert K)=
\frac{\Gamma(\alpha)\prod_{k=1}^K \Gamma(n_k+\alpha/K)}{\Gamma(\alpha/K)^K\Gamma(n+\alpha)}. \label{allo}
\end{equation}

The distributional results presented in \eqref{post_mu}, \eqref{marginal_left}, and \eqref{allo} \citep[for a proof, see, for example,][]{murphy} are the building blocks for the MCMC sampler which is used to make Bayesian inference on the model parameters of interest.

\section{Inference via MCMC} \label{mcmc_inference}

Since the full posterior is not analytically tractable, inference is performed using MCMC sampling with trans-dimensional moves \citep{Green95}. The main objective of the analysis is to cluster the nodes, and therefore the locations $\vec \mu_k$, the variance parameters $\vec\Sigma_k$ and $\bm\sigma^2_k$ and the community probabilities $\bm\theta$ are considered as nuisance parameters and integrated out. Essentially, in this type of collapsed Gibbs sampler \citep{liu}, four moves are available \citep{richardson,zhang}, described in the subsequent four subsections.

\subsection{Change in the community allocations} \label{change_community}

A fully collapsed Gibbs update for each community assignment is available:
\begin{equation}
  p(z_i=k\vert\vec z_{-i},{\mvec X},d,K) \propto p(z_i=k\vert\vec z_{-i},d,K) p({\vec x}_i\vert \{{\vec x}_j\}_{j\neq i:z_j=k},d).
  \label{comm_resamp}
\end{equation}
In the special case where $d=K_\varnothing$ and $n_{z_i}=1$, the full conditional distribution for $z_i$ assigns probability one to retaining the same value since the model does not permit $d>K_\varnothing$. Otherwise, from \eqref{allo}:
\begin{equation}
p(z_i=k\vert\vec z_{-i},d,K) \propto 
\frac{n_k^{-i} + \alpha/K}{n-1+\alpha}. \label{popularity}
\end{equation}
where $n_k^{-i}=n_k-\mathds 1_k(z_i)$. Similarly, the remaining term in \eqref{comm_resamp}, $p({\vec x}_i\vert \{{\vec x}_j\}_{j\neq i:z_j=k},d)$, can be obtained as the ratio of marginal likelihoods 
\begin{equation}
p({\vec x}_i\vert \{{\vec x}_j\}_{j\neq i:z_j=k}, d) = \frac{p({\vec x}_i,\{{\vec x}_j\}_{j\neq i:z_j=k}\vert d)}{p(\{{\vec x}_j\}_{j\neq i:z_j=k}\vert d)}. \label{mar_lik_ratio}
\end{equation}
The ratio \eqref{mar_lik_ratio} can be decomposed as the product of two ratios of marginal likelihoods. Using \eqref{marginal_left}, 
 the first ratio is equivalent to the following multivariate Student $t$ distribution \citep{murphy}:
\begin{equation}
p(\xid\vert \{\xjd\}_{j\neq i:z_j=k},d) = t_{\nu_{n_k^{-i}}}\left(\xid\left\vert\m,\frac{\kappa_{n_k^{-i}}+1}{\kappa_{n_k^{-i}}\nu_{n_k^{-i}}} \Dki\right)\right. .
\label{t_left}
\end{equation}
where the additional superscript $-i$ denotes a cluster quantity that is computed excluding the allocation $z_i$ of the $i$-th node.
The second ratio, which accounts for the last $m-d$ dimensions, has the form
\begin{equation}
p(\xidd\vert \{\xjdd\}_{j\neq i:z_j=k},d) = \prod_{j=d+1}^m t_{\lambda_{n_k^{-i}}}\left(x_{ij}\left\vert 0,\ski\right)\right. . \label{t_prod}
\end{equation}

\subsection{Split or merge two communities} \label{split_merge}

To vary the number of communities, move proposals inspired by Sequentially-Allocated Merge-Split sampling \citep{dahl} are used here, but other alternative choices are available \citep[see, for example,][]{jain,particle}. Two indices $i$ and $j$ are sampled at random from the $n$ nodes, and without loss of generality assume $z_i\leq z_j$. If $z_i=z_j$, then the single cluster is split, whereas if $z_i> z_j$ the two clusters are merged. In both move types, node $i$ will remain in the same cluster, denoted $z^\star_i=z_i$. In the merge move, all elements of cluster $z_j$ are reassigned to cluster $z_i$ (with any higher indexed clusters subsequently decremented). For the split move, node $j$ is first reassigned to cluster $K^\star=K+1$ with new allocation $z^\star_{j}=K^\star$; then, in random order the remaining nodes currently allocated to cluster $z_i$ are randomly reassigned to clusters $z_i$ or $K^\star$ with probability proportional to their predictive distribution from the generative model \eqref{mar_lik_ratio}. Denoting the resulting product of renormalised predictive densities from these reallocations by $q(K^\star,\vec z^\star\vert K, \vec z)$, the acceptance probability for a split move, for example, is
\begin{equation}
\alpha(K^\star,\vec z^\star\vert K,\vec z) = \min\left\{ 1, \frac{p(\mvec X\vert d,K^\star,\vec z^\star)
p(d\vert\vec z^\star,K^\star)p(\vec z^\star\vert K^\star)p(K^\star)}{p(\mvec X\vert d,K,\vec z)
  p(d\vert\vec z,K)p(\vec z\vert K)p(K)q(K^\star,\vec z^\star\vert K, \vec z)}
\right\}.
\label{accept_ratio_split_merge}
\end{equation}
Note that the ratio of densities for $\mvec X$ in the acceptance ratio \eqref{accept_ratio_split_merge} will depend only upon the rows of the matrix corresponding to the cluster being split (or similarly, merged), and furthemore these expressions will decompose as a products of terms for the first $d$ and remaning $m-d$ components (\textit{cf.} \eqref{t_left}, \eqref{t_prod}).

\subsection{Create or remove an empty community} \label{propose_empty}

Adding or removing empty communities whilst fixing $\vec z$ corresponds to proposing $K^\star=K+1$ or $K^\star=K-1$ respectively, although the latter proposal is not possible if $K=K_\varnothing$, meaning there are currently no empty communities. 
The acceptance probability is simply
\begin{equation}
  \alpha(K^\star\vert K )= \min\left\{ 1, \frac{p(\vec z\vert K^\star)p(K^\star)q_\varnothing}{p(\vec z\vert K)p(K)}
  \right\},
\end{equation}
where the proposal ratio $q_\varnothing=2$ if $K^\star=K_\varnothing$, $q_\varnothing=0.5$ if $K=K_\varnothing$ and $q_\varnothing=1$ otherwise.


\subsection{Change in the latent dimension} \label{change_dimension}
This move is only required when the latent dimension is not marginalised out. Given a current value $d$, a new value $d^\star$ is proposed from a density $q(d^\star\vert d)\propto \xi^{\abs{d^\star-d}}\mathds 1_\mathcal{D}(d^\star)$ on a neighbourhood
\begin{equation}
  \mathcal{D}=\{\max\{1,d-l\},\dots,d-1,d+1,\dots,\min\{d+l,m\}\},
\end{equation}
typically with $l\leq 5$ and with $\xi\in(0,1)$. The acceptance ratio reduces to

\begin{equation}
\alpha(d^\star\vert d) = \min\left\{ 1, \frac{p(\mvec X\vert d^\star,K,\vec z)p(d^\star\vert\vec z)}{p(\mvec X\vert d,K,\vec z)p(d\vert\vec z)}\frac{q(d\vert d^\star)}{q(d^\star\vert d)} \right\}.
\end{equation}

Notably, if $d^\star>d$, the ratio $p(\mvec X\vert d^\star,K,\vec z)/p(\mvec X\vert d,K,\vec z)$ only depends on the first $d^\star$ components of the embedding, since the last $m-d^\star$ components remain independent by \eqref{full_model}.

\subsection{Inferring communities}\label{sec:infer_communities}
Markov Chain Monte Carlo samplers for mixture models with varying number of clusters are well known to be affected by \textit{label switching} \citep{jasra}, since the likelihood is invariant to permutations of the cluster labels. However, 
the estimated posterior similarity between nodes $i$ and $j$, $\hat\pi_{ij}=\hat{\mathbb P}(z_i=z_j\vert\mvec X)=\sum_{s=1}^{M} \mathds 1_{z_i^{(s)}}\{z_j^{(s)}\}/M$ is invariant to label switching.
Communities can be estimated from the MCMC chains using the posterior similarity matrix and the PEAR method \citep[maximisation of the posterior expected Rand adjusted index,][\texttt{R} package \texttt{mcclust}]{fritsch}. 
Alternatively, if a configuration with a fixed number of clusters $K$ is required, the clusters can been estimated using hierarchical clustering with average linkage, using $1-\hat\pi_{ij}$ as distance measure \citep{medve}. 
Many alternatives to this method have been proposed in the literature \citep{rand,binder,dahl_clust}.

\section{Second-level clustering of community variances} \label{curse_section}

Empirical analysis of assuming the model \eqref{full_model} for simulations from the stochastic block model show that identifying and clearly separating the $K$ clusters in $\mvec X_{d:}$ is particularly difficult for the sampler in settings when $d\ll m$. The problem is particularly evident when $m=n$ and $d$ is small. In this case, it has been assessed empirically that the within-cluster variance of the true communities in simulated datasets seems to converge to similar values, such that $\sigma^2_{k,j}\approx\sigma^2_{\ell ,j}$ for $j\gg d$ and $k\neq \ell$. Therefore, when $m$ is large enough, the selected model tends to be under-specified: the correct dimension $d$ is identified, but the true number of communities $K$ is underestimated. 
This is also one of the main reasons why it is not advisable to directly fit a Gaussian mixture model on $\mvec X\in\mathbb R^{n\times m}$ and allow $K$ to be random, ignoring the role of $d$. 

\begin{figure}[!b]
  \centering
  \includegraphics[]{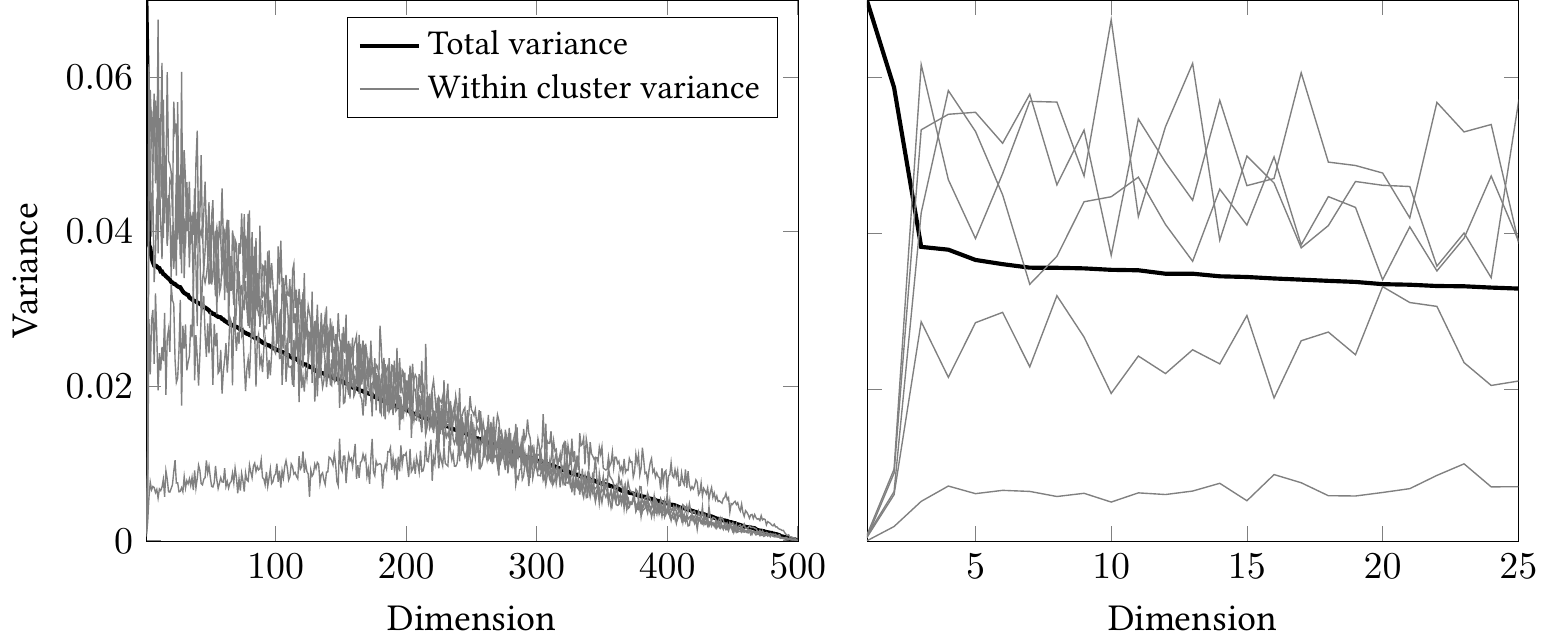}

\caption{Within-block variance and total variance for the adjacency embedding obtained from a simulated $5$-block SBM with $d=2$, $n=500$, and well separated means $\bm\mu_1=[0.7,0.4], \bm\mu_2=[0.1,0.1], \bm\mu_3=[0.4,0.8], \bm\mu_4=[-0.1,0.5]$ and $\bm\mu_5=[0.3,0.5]$, and $\bm\theta=(0.2,0.2,0.2,0.2,0.2)$. The right panel is the left panel plot zoomed in to the first 25 dimensions.}
\label{fig:curse}
\end{figure}

The problem is illustrated in Figure~\ref{fig:curse}, which shows the results from performing ASE for a simulation of $n=500$ nodes from a stochastic block model containing five equally probable communities with well-separated mean locations. The within-cluster variance of three of the five communities fluctuate around the same values on each dimension larger than $d$. For a dimension larger than approximately $150$, four of the five clusters have the same variance on the subsequent dimensions. Therefore, when $m=n$, the MCMC sampler selects the MAP estimate $\hat K=2$ for parsimony, and increases the variance of the Gaussian distributions on the first two dimensions, on which the clusters are well separated.

The solution proposed here is to assume shared variance parameters between some of the clusters for dimensions larger than $d$. Specifically, each community $k\in\{1,\dots,K\}$ is assigned a second-level cluster allocation $v_k\in\{1,\dots,H\}$, with $H\leq K$. If $v_k=v_{\ell}$, then for $j>d$, $\sigma^2_{k,j}=\sigma^2_{\ell,j}$. Formally,
\begin{align}
\vec x_i \vert d,K,z_i,v_{z_i},\vec \mu_{z_i},\bm\Sigma_{z_i}, \bm\sigma^2_{v_{z_i}} &\overset{d}{\sim} \mathbb N_m \left( \begin{bmatrix} \vec \mu_{z_i} \\ \vec 0 \end{bmatrix}, \begin{bmatrix} \bm\Sigma_{z_i} & \vec 0 \\ \vec 0 & \bm\sigma^2_{v_{z_i}}\mvec I_{m-d} \end{bmatrix} \right),\ i=1,\dots,n, \\
v_k \vert K,H &\overset{d}{\sim}\mathrm{Multinoulli}(\bm\phi),\ k=1,\dots,K, \\
\bm\phi\vert H &\overset{d}{\sim} \mathrm{Dirichlet}\left({\beta}/{H},\dots,{\beta}/{H}\right), \\
H\vert K&\overset{d}{\sim} \mathrm{Uniform}\{1,\dots,K\}. 
\end{align}

\begin{figure}[!t]
\centering
\includegraphics[width=.7\textwidth]{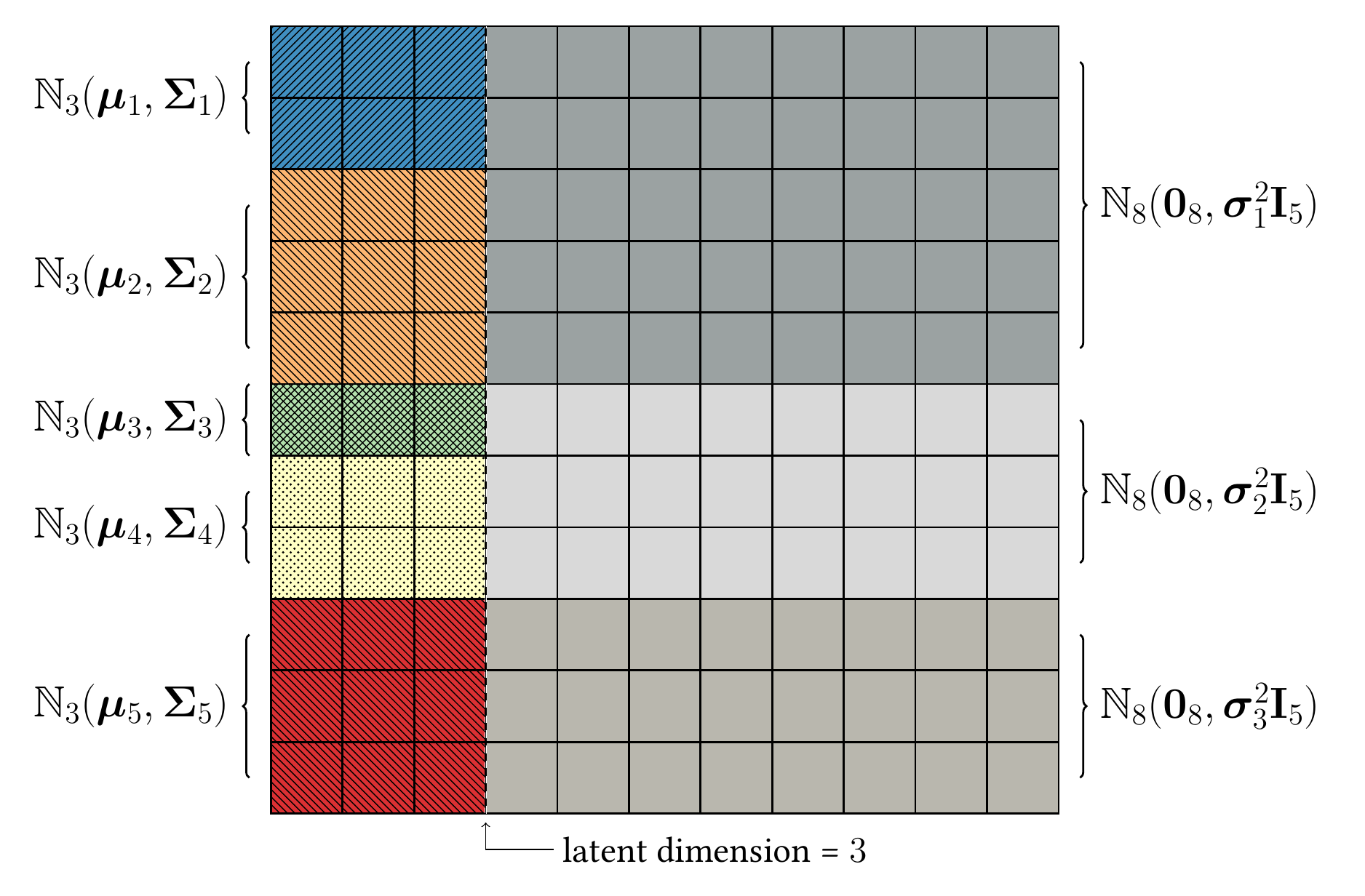}
\caption{Cartoon for the generating process of the embedding of an 11-node stochastic blockmodel GRDPG with $K=5$ communities and latent dimension $d=3$, with common variance for communities 1-2 and 3-4 on the right hand side of the matrix.}
\label{grid_curse}
\end{figure}

Essentially, the vector $\vec v=(v_1,\dots,v_K)$ defines a \textit{clustering of communities}. Figure~\ref{grid_curse} depicts a cartoon example of this extended model. Note that if $H=1$ and all the communities are assigned to the same second-level cluster, the problem of selecting $d$ essentially reduces to an \textit{ordinal} version of the feature selection problem in clustering \citep{dean,maugis}.



Under this extended model, the posterior distribution for $\sigma^2_{j,k}$ changes due to the aggregation of communities in the second level. Under the $\mathrm{Inv}\text{-}\chi^2(\lambda_0,\sigma^2_0)$ prior, the posterior is $\mathrm{Inv}\text{-}\chi^2(\lambda_{n_{\bullet k}},\sdotkj)$ where
\begin{equation}
n_{\bullet k} = \sum\nolimits_{\ell:v_\ell=k} n_{\ell}, \quad \sdotkj = \left\{\lambda_0\sigma^2_{0} + \sum\nolimits_{i:v_{z_i}=k} x_{ij}^2\right\}\bigg/\lambda_{n_{\bullet k}}. \label{eq:sigma_post_pars2}
\end{equation}

Calculations similar to \eqref{marginal_right} give the correct form of the marginal likelihood for the right hand side of the matrix, restricted to a given value of $v_k$. Clearly, $\bm\phi$ can be again marginalised out, yielding the marginal likelihood
\begin{equation}
p(\vec v\vert H) = \frac{\Gamma(\beta)\prod_{h=1}^H \Gamma(\sum_{k=1}^K \mathds 1_{h}\{v_k\}+\beta/H)}{\Gamma(\beta/H)^H\Gamma(K+\beta)}.
\end{equation}

The MCMC sampler described in Section \ref{mcmc_inference} must be slightly adapted. For the Gibbs sampling move in Section~\ref{change_community}, the product of univariate Student's $t$ densities in \eqref{t_prod} is modified using the appropriate $(\lambda_{n_{\bullet k}},\sdotkj)$
 pair. For the change in dimension, the likelihood $p(\mvec X\vert d,K,\vec z, \vec v)$ should be computed using the shared variances and the allocations $\vec v$. When an empty community is proposed, as in Section~\ref{propose_empty}, the ratio $p(\vec v^\star\vert K)/p(\vec v\vert K)$ must be added, limited to the second level allocation of the additional community. The value $v_k$ for the proposed empty cluster can be simply chosen at random from $\{1,\dots,H\}$. Finally, for the split-merge move in Section~\ref{split_merge}, if $z_i=z_j$ for the two selected nodes, then $v_{z_i}=v_{z_j}$ after the split move. Alternatively, if $z_i\neq z_j$, then the new value of $v_k$ is sampled at random from $v_{z_i}$ and $v_{z_j}$.  

Finally, three additional moves are required: resampling the second-level cluster allocations $\vec v$ using a Gibbs sampling step; proposing a second-level split-merge move; and adding or removing an empty second-level cluster. When $\bm\phi$ and the parameters of the Gaussian distributions are marginalised out, the second-level allocations are resampled according to the following equation:
\begin{equation}
p(v_k=h\vert\vec v_{-k},{\mvec X},\vec z,d,K) \propto p(v_k=h\vert\vec v_{-k},K) p(\Xkdd\vert \{\mvec X^{(\ell)}_{d:}\}_{\ell\neq k:v_\ell=h},v_k=h,d,K), \label{second_order_resamp}
\end{equation}
where the independence assumption between $\Xkd$ and $\Xkdd$ is used. Similarly to \eqref{popularity}:
\begin{equation}
p(v_k=h\vert\vec v_{-k},K) = \frac{\sum_{\ell\neq k} \mathds 1_h\{v_\ell\}+\beta/H}{K-1+\beta}.
\end{equation}
The calculations for the second term in \eqref{second_order_resamp} are similar to \eqref{mar_lik_ratio}:
\begin{equation}
p(\Xkdd\vert \{\mvec X^{(\ell)}_{d:}\}_{\ell\neq k:v_\ell=h},v_k=h,d,K) = 
\frac{p(\Xkdd,\{\mvec X^{(\ell)}_{d:}\}_{\ell\neq k:v_\ell=h}\vert v_k=h,d,K)}{p(\{\mvec X^{(\ell)}_{d:}\}_{\ell\neq k:v_\ell=h}\vert d,K)},
\end{equation}
which can be computed using \eqref{marginal_right}. 
The second-level split-merge move and the proposal of an empty cluster follows the same guidelines in Sections~\ref{split_merge} and~\ref{propose_empty}.

Potentially, the model could be extended further using the same reasoning: from the plot in Figure~\ref{fig:curse}, it is clear that the different clusters begin to share the same variance at different points in the plot. Empirically, all the variances approximately converge to the same values at large dimensions, and it is therefore possible to identify a $(K-1)$-vector of discrete points in $\{d,d+1,\dots,m\}$ at which different community variances coalesce. 
For the plot in Figure~\ref{grid_curse}, such vector could be $(d,d,d,150,n)$, with $d=2$ and $n=m=500$.

\section{Extension to directed and bipartite graphs} \label{ext}


A directed graph $\mathbb G=(V,E)$ has the property that $(i,j)\in E \centernot\implies (j,i)\in E$, meaning the corresponding adjacency matrix $\mvec A\in\{0,1\}^{n\times n}$ is not, in general, symmetric. Directed graphs are useful for representing directed interaction networks, such as email traffic patterns; knowing that individual $i$ broadcasts emails to individual $j$ does not immediately imply that $j$ also issues communications to $i$.

\begin{definition}[Adjacency embedding of the directed graph] \label{adj_emb_dir}
Given a directed graph with adjacency matrix $\mvec A\in\{0,1\}^{n\times n}$, and a positive integer $d$, $1\leq d\leq n$, consider the singular value decomposition 
\begin{equation}
\mvec A = \begin{bmatrix} \hat{\mvec U} & \hat{\mvec U}_\perp\end{bmatrix} \begin{bmatrix} \hat{\mvec D} & \mvec 0 \\ \mvec 0 & \hat{\mvec D}_\perp \end{bmatrix} \begin{bmatrix} \hat{\mvec V}^\top \\ \hat{\mvec V}^\top_\perp \end{bmatrix} = \hat{\mvec U}\hat{\mvec D}\hat{\mvec V}^\top + \hat{\mvec U}_\perp\hat{\mvec D}_\perp\hat{\mvec V}^\top_\perp,
\end{equation}
where $\hat{\mvec D}\in\mathbb R_+^{d\times d}$ is diagonal matrix containing the top $d$ singular values in decreasing order, $\hat{\mvec U}\in\mathbb R^{n\times d}$ and $\hat{\mvec V}\in\mathbb R^{n\times d}$ contain the corresponding left and right singular vectors, and the matrices $\hat{\mvec D}_\perp$, $\hat{\mvec U}_\perp$, and $\hat{\mvec V}_\perp$ contain the remaining $n-d$ singular values and vectors. The $d$-dimensional directed adjacency embedding of $\mvec A$ in $\mathbb R^{d}$, is defined as the pair
\begin{equation}
  \hat{\mvec X}=\hat{\mvec U}\hat{\mvec D}^{1/2}, \quad \hat{\mvec X^\prime}=\hat{\mvec V}\hat{\mvec D}^{1/2}.
\end{equation}
\end{definition}


Writing $\Xs={\mvec U}{\mvec D}^{1/2}$ and $\Xr={\mvec V}{\mvec D}^{1/2}$, the rows of $\Xs$ characterise the activities of each node as a source, and the rows of $\Xr$ characterise the same nodes as destinations.


The model in \eqref{full_model} can be easily adapted to directed graphs. Treating the embeddings $\Xs$ and $\Xr$ as independent, each is modelled separately using the same Gaussian structure and prior distributions \eqref{full_model}, except for three parameters which are initially assumed common to both embeddings: the latent dimension $d$, the number of communities $K$ and the vector of node assignments to those communities, $\vec z$.

In some contexts it will be more relevant to consider different community membership structures for the same set of nodes when considering them as source nodes or destination nodes. In this case, let $K$ denote the number of source communities and $K^\prime$ denote the number of destination communities; similarly let $\vec z$ denote the assignments of nodes to source communities, and $\vec z^\prime$ the allocations to destination communities. Jointly learning $\vec z$ and $\vec z^\prime$ (as well as $d$) is a problem commonly known as \textit{co-clustering}, and the corresponding network model is known as the stochastic co-blockmodel (ScBM) \citep{scbm}. 
Given an asymmetric matrix $\mvec B\in[0,1]^{K\times K^\prime}$, then 
$\mathbb P(A_{ij}=1)=B_{z_iz^\prime_j}$. 
From a random dot product graph perspective, it is assumed that 
$B_{z_iz^\prime_j}=\bm\mu_{z_i}^\top\bm\mu^\prime_{z^\prime_i}$, 
for some latent positions $\bm\mu_{z_i},\bm\mu^\prime_{z^\prime_i}\in\mathbb R^d$
and $d=\rank(\mvec B)\leq\min(K,K^\prime)$.

The Bayesian model for ScBMs can be easily represented as a separate model for $\Xs$ and $\Xr$, of the form given in \eqref{full_model}, with the latent dimension of the embedding $d$ now the only common parameter. 
Inference via MCMC can be performed in an equivalent way to the method described in Section~\ref{mcmc_inference}; the only difference is in the expression of the acceptance ratio for a change in the shared latent dimension $d$, but the procedure can exploit the results obtained in Section~\ref{change_dimension}, using the fact that
\begin{equation}
p(\mvec X,\mvec X^\prime\vert d,K,K^\prime,\vec z,\vec z^\prime) =\prod_{k=1}^{K} p(\mvec X^{(k)}_{:d}\vert d)p(\mvec X^{(k)}_{d:}\vert d)\prod_{k^\prime=1}^{K^\prime} p(\mvec {X^{\prime}}^{(k^\prime)}_{:d}\vert d)p(\mvec {X^{\prime}}^{(k^\prime)}_{d:}\vert d),
\end{equation}
where all the marginal likelihoods can be equivalently obtained from \eqref{marginal_left}. 
Furthermore, the model 
can be appropriately modified when $d\ll m$ to include the second-level cluster allocations proposed in Section~\ref{curse_section}.


Finally, in bipartite graphs, the observed nodes can be partitioned into two sets $V$ and $V^\prime$, with $V\cap V^\prime=\varnothing$ and $E\cap(V\times V \cup V^\prime\times V^\prime)=\varnothing$. Assume that $V$ plays the role of the set of source nodes and $V^\prime$ of the set of destination nodes. Bipartite graphs are usually represented by a rectangular bi-adjacency matrix $\mvec A\in\{0,1\}^{n\times n^\prime}$, with $n^\prime=\abs{V^\prime}$. In this case, it is still possible to apply the methods described in this section to the SVD embedding obtained from the rectangular matrix $\mvec A$. Note that the ScBM 
extends trivially to the bipartite graph case, which is essentially a special case of a directed graph, with the cluster configurations for source and destination nodes now inescapably unrelated and each node possessing only one latent representation in $\mathbb{R}^d$.


\section{Applications and results} \label{results}

The Bayesian latent feature network models described in this article have been applied to a both simulated and real world network data from undirected, directed and bipartite graphs.
 The real network data analysed are from an undirected network obtained from the Santander bikes hires in London, and the Enron Email Dataset; details are given in the corresponding sections.

The model and MCMC sampler have been tested using different combinations of the hyperparameters, showing robustness to the prior choice. Inferential performance is sensitive to extreme values of the variance parameters, relative to their prior mean, but otherwise robust. So in practice, the expectation of the prior for the variance parameters should be chosen to be on the same scale as the observed data. 
The default settings for the MCMC sampler used in the next sections are the usual \textit{uniformative} values $\kappa_0=\nu_0=\lambda_0=\alpha=\beta=1$, and $\omega=\delta=0.1$. For the proposal of change in dimension (\textit{cf.} Section \ref{change_dimension}), $\xi=0.8$.  
The algorithms were run for a total of $M=\numprint{500000}$ samples with burn-in $\numprint{25000}$, for a number of different chains to check for convergence. 
The cluster configuration has been initialised using $K$-means for some pre-specified $K$, usually chosen according to the scree-plot criterion. 
The second-level clusters have been initialised setting $H=K$. 
In order to set the prior covariances to a realistic value, the correlations in the $\bm\Delta_d$ matrices are set to zero, and the elements on the diagonal of $\bm\Delta_d$ to the average within-cluster variance based on the $K$-means cluster configuration.
Similarly, the prior $\sigma^2_{0j}$ values have been set to the total variance on the corresponding column of the embedding. 


\subsection{Synthetic data and model validation}  \label{section_vali}

A stochastic blockmodel can be simulated starting from a matrix $\mvec B\in[0,1]^{K\times K^\prime}$ containing the probabilities of connection between communities, and a vector $\bm\theta$ of community allocation probabilities. Here, each element $B_{k\ell}$ of $\mvec B$ was generated from a $\mathrm{Beta}(1.2,1.2)$ distribution, which produces communities with a moderate level of separation. For an undirected graph, $K=K^\prime$ and the constraint $B_{k\ell}=B_{\ell k}$ is imposed; similarly, for directed graphs with a shared cluster configuration (\textit{cf.} Section \ref{ext}), $K=K^\prime$.

The latent dimension $d$ can be interpreted as the rank of the matrix $\mvec B$, and a random matrix $\mvec B$ generated from independent beta draws has full rank with probability $1$. Therefore, to simulate $d<K$ a low-rank approximation of $\mvec B$ must be used to generate the embedding. 
For undirected graphs, a truncated spectral decomposition can be used: $\tilde{\mvec B}=\bm\Gamma_d\bm\Lambda_d\bm\Gamma_d^\top$ (recall Definition~\ref{adj_emb}). Similarly, for the directed and bipartite graphs, the truncated SVD is an appropriate approximation: $\tilde{\mvec B}= \mvec U_d\mvec D_d\mvec V_d^\top$ (see Definition~\ref{adj_emb_dir}). Note that under this low-rank approximation, it must be checked that each  element $\tilde{B}_{k\ell}\in [0,1]$. 

\begin{figure}[p]
\centering

\begin{subfigure}[t]{0.48\textwidth}
\centering
\caption{Scatterplot of the simulated $\mvec X_1$ and $\mvec X_2$}
\includegraphics[width=.94\textwidth]{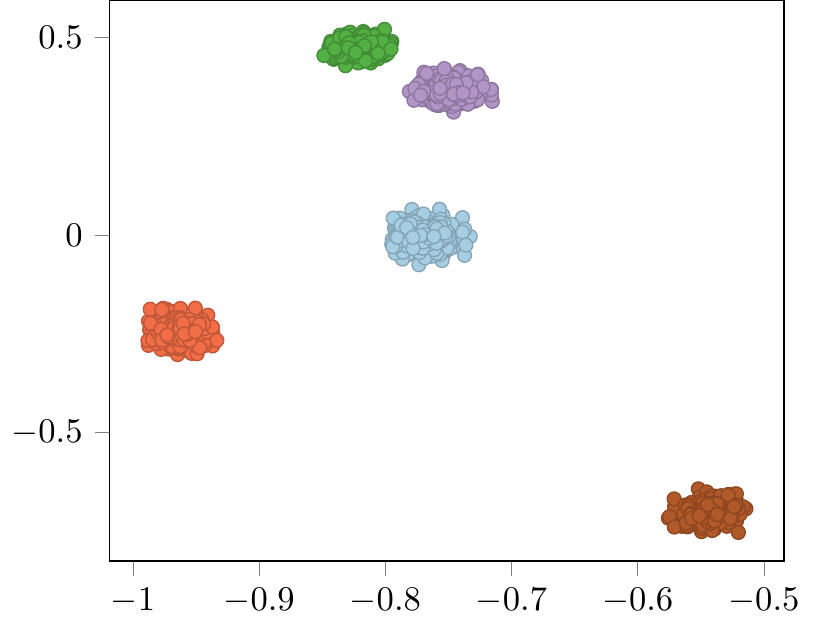}
\label{scatter_sim}
\end{subfigure}
\begin{subfigure}[t]{0.48\textwidth}
\centering
\caption{Scatterplot of the simulated $\mvec X_3$ and $\mvec X_4$}
\includegraphics[width=.94\textwidth]{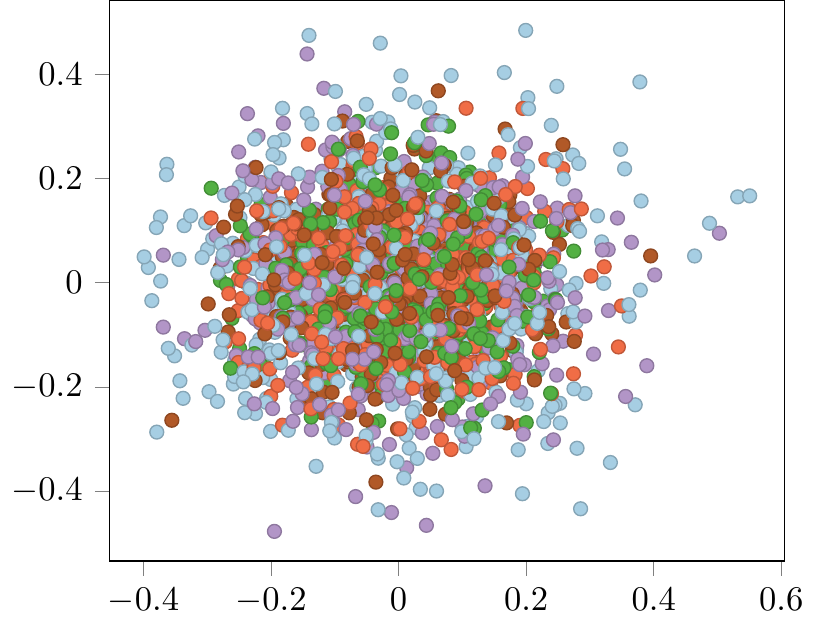}
\label{scatter_sim_rest}
\end{subfigure}

\begin{subfigure}[t]{0.48\textwidth}
\centering
\caption{Within-cluster and overall means of $\mvec X_{:15}$}
\includegraphics[width=.93\textwidth]{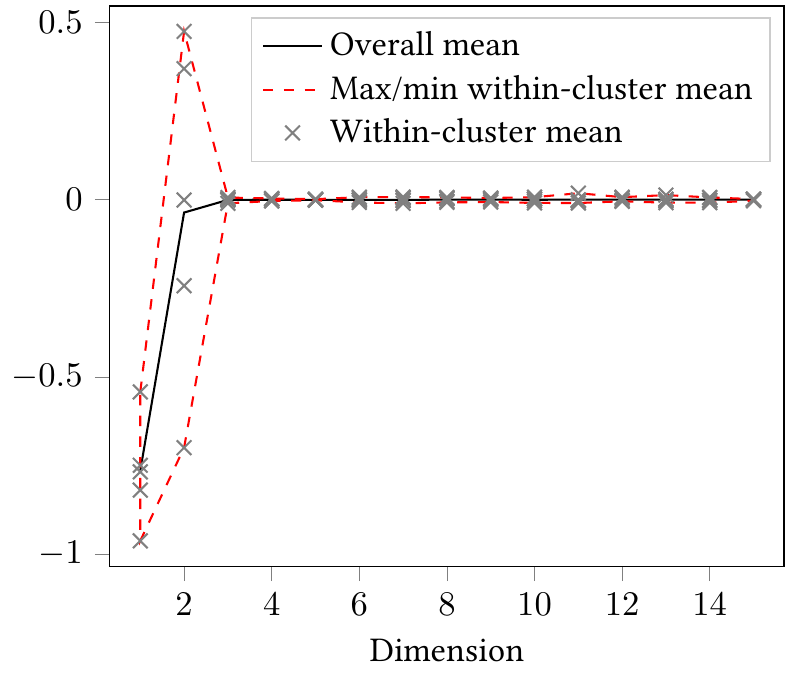}
\label{mean_und}
\end{subfigure}
\begin{subfigure}[t]{0.48\textwidth}
\caption{Within-cluster and overall variances of $\mvec X_{:25}$}
\includegraphics[width=.98\textwidth]{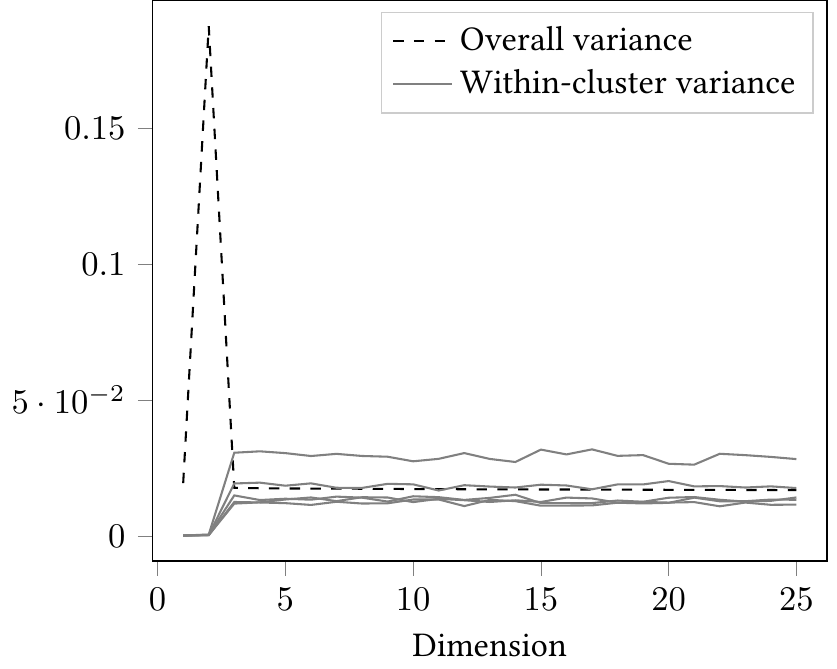}
\label{var_und}
\end{subfigure}

\begin{subfigure}[t]{0.48\textwidth}
\centering
\caption{Within-cluster correlation coefficients of $\mvec X_{:30}$}
\includegraphics[width=.98\textwidth]{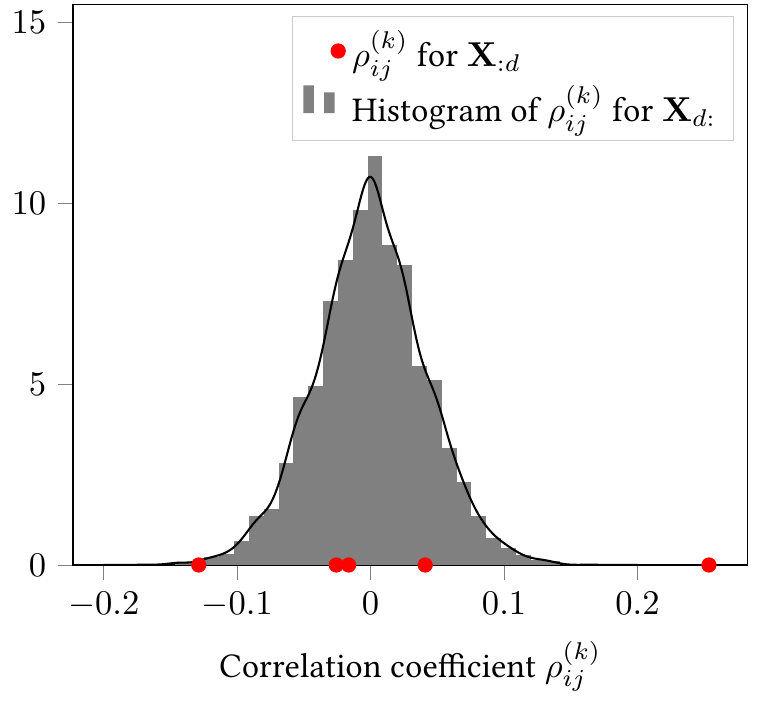}
\label{corr_und}
\end{subfigure}
\begin{subfigure}[t]{0.48\textwidth}
\centering
\caption{Marginal likelihood as function of $d$ (\textit{true} clusters)}
\includegraphics[width=.92\textwidth]{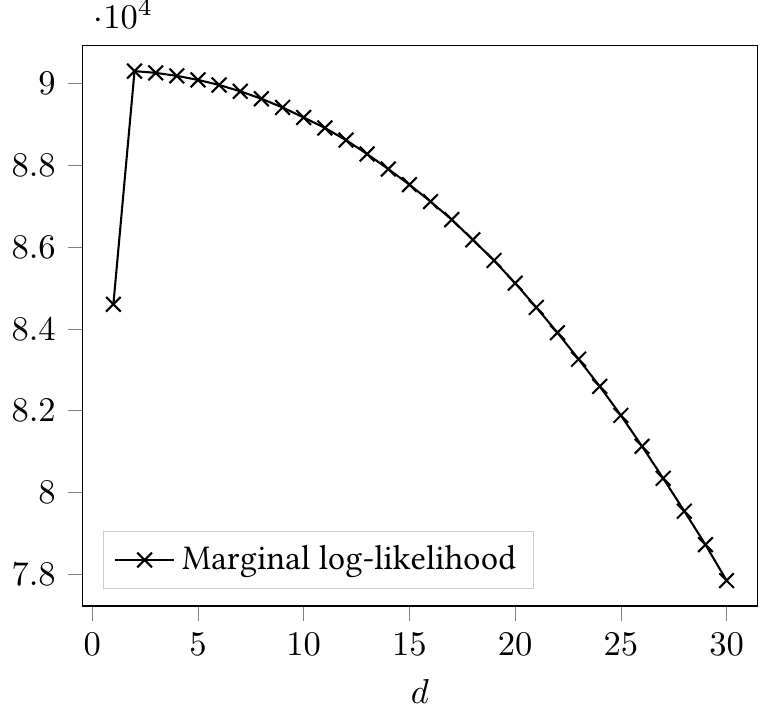}
\label{mlik_und}
\end{subfigure}

\caption{Adjacency embedding for an undirected graph with $n=2500$ nodes, $K=5$, obtained from a symmetric $\mvec B\in[0,1]^{K\times K}$ with $B_{k\ell}\overset{d}{\sim}\mathrm{Beta}(1.2,1.2)$, $d=2$ and $m=50$.}
\label{undirected_simulated}
\end{figure}

Figure~\ref{undirected_simulated} shows results for a synthetic undirected graph with $d=2$ and $K=5$. The scatterplot of the first two columns of the adjacency embedding $\mvec X$, plotted in Figure~\ref{scatter_sim}, show well-separated clusters, which can be suitably modelled using a Gaussian mixture. Figure~\ref{scatter_sim_rest}, shows the next two dimensions. Clearly, the community mean locations are significantly different from zero in just the first two dimensions, and this is further illustrated in Figure~\ref{mean_und}. Similarly, in Figure~\ref{var_und}, the difference between the within-cluster and overall variance is evident only in the first two dimensions, after which the quantities are of the same order of magnitude. The within-cluster variances differ across communities, suggesting that it is appropriate to have cluster-specific values of $\sigma^2_{j,k}$ for $j>d$; this phenomenon can also be witnessed in Figure~\ref{scatter_sim_rest}. Nevertheless, it also seems appropriate to use a second-level clustering with $H=3$, since the variances of three of the five communities are approximately the same for dimensions larger than $d=2$. Also, in Figure~\ref{corr_und}, the within-cluster correlations between $\mvec X_1$ and $\mvec X_2$ suggest dependence for at least one of the clusters. On the other hand, the sample within-cluster correlations for $\Xdd$ tend to be small and centred around 0, suggesting that the assumption of independence is appropriate in that part of the model \eqref{full_model}. Finally, form Figure~\ref{mlik_und} the marginal likelihood strongly favours the true value $d=2$, resulting in a posterior distribution essentially consisting of a point mass at the true value.

\begin{figure}[!t]
\centering

\begin{subfigure}[t]{0.48\textwidth}
\centering
\caption{Scatterplot of the simulated $\mvec X_{:d}$}
\includegraphics[width=1.02\textwidth]{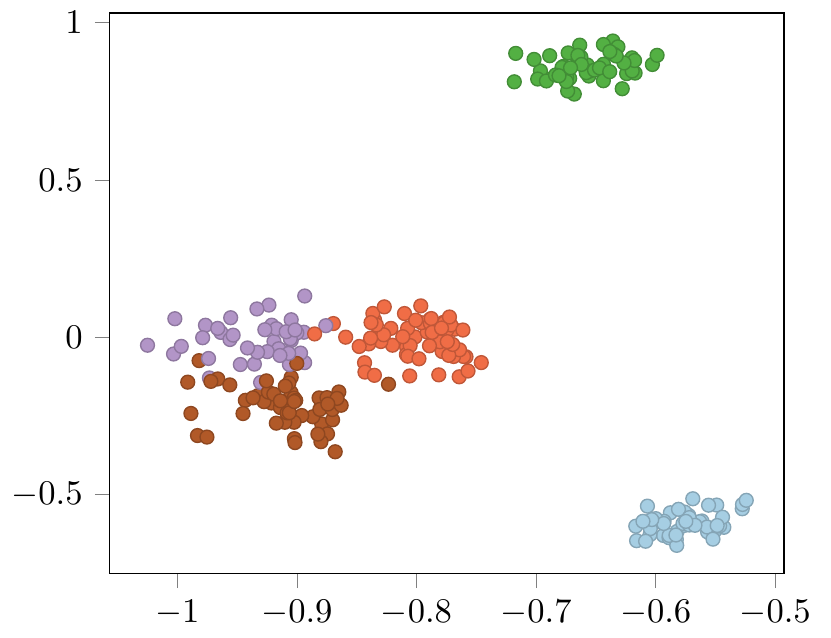}
\label{scatter_s}
\end{subfigure}
\begin{subfigure}[t]{0.48\textwidth}
\centering
\caption{Scatterplot of the simulated $\mvec X^\prime_{:d}$}
\includegraphics[width=.99\textwidth]{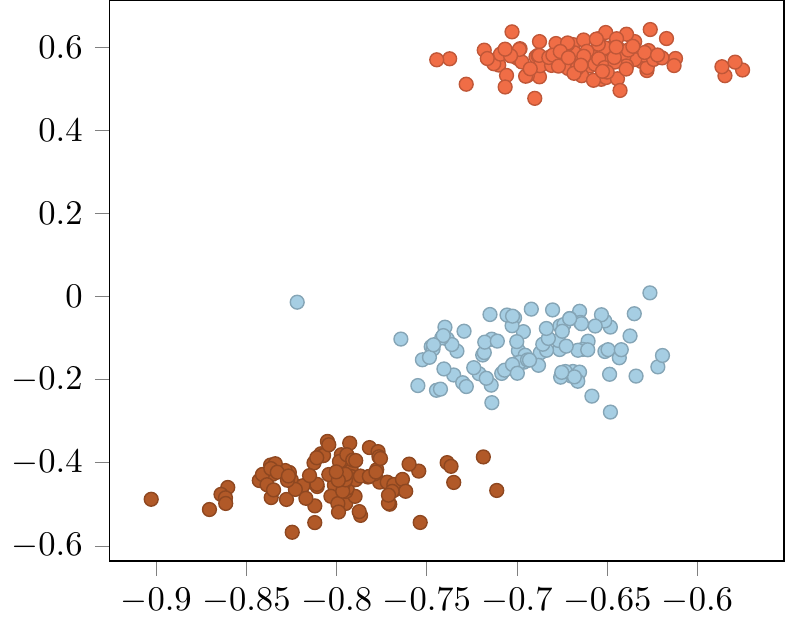}
\label{scatter_r}
\end{subfigure}

\begin{subfigure}[t]{0.48\textwidth}
\centering
\caption{Within-cluster and overall means of $\mvec X_{:15}$}
\includegraphics[width=\textwidth]{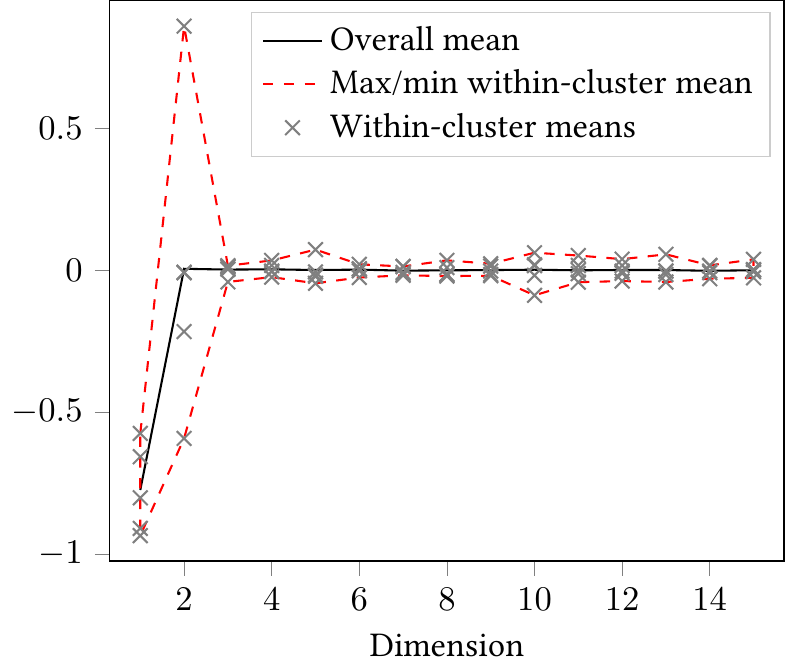}
\label{mean_sim}
\end{subfigure}
\begin{subfigure}[t]{0.48\textwidth}
\centering
\caption{Marginal likelihood as function of $d$ (\textit{true} clusters)}
\includegraphics[width=\textwidth]{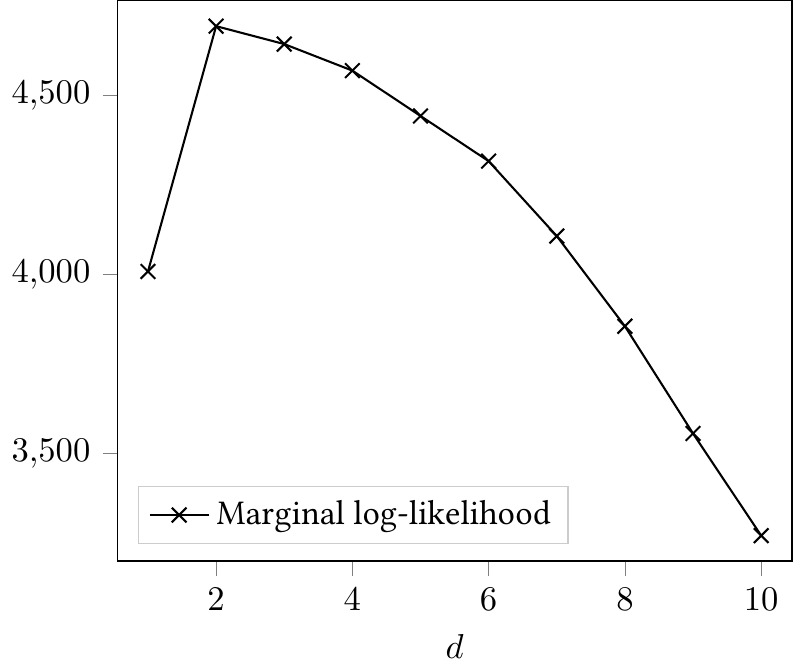}
\label{mlik_bip}
\end{subfigure}

\caption{Simulated adjacency embedding for a bipartite $250\times300$ graph with $K=5$ and $K^\prime=3$, obtained from $\mvec B\in[0,1]^{K\times K^\prime}$ with $B_{k\ell}\overset{d}{\sim}\mathrm{Beta}(1.2,1.2)$ and $d=2$.}
\label{bipartite_simulated}
\end{figure}

Figure~\ref{bipartite_simulated} shows results for a simulated bipartite graph with separate community structures for nodes as sources and destinations, with $d=2$, $K=5$ and $K^\prime=3$. Again, the clusters in Figure~\ref{scatter_s} and~\ref{scatter_r} are well-separated and can be easily estimated using the Gaussian mixture model, both for sources and receivers. From Figure~\ref{mean_sim}, the zero-mean assumption for the columns with index larger than $d$ seems to hold even for a relatively small number of nodes per community. A similar plot can be produced for $\Xr$, showing a similar pattern. In Figure~\ref{mlik_bip}, the marginal likelihood strongly favours the true value $d=2$, which again results in a point mass posterior centred at the true value. Similar considerations hold for variances and correlations, with results which are similar to the plots in Figure~\ref{var_und} and~\ref{corr_und} for the undirected graph.

Overall, for synthetic data, the model seems robust and able to detect the correct $d$ and $K$ in a variety of different settings. 

\subsection{Undirected graphs: Santander bikes}

The \textit{Santander Cycle hire scheme} is a bike sharing system implemented in central London. Transport for London periodically releases data on the bike hires\footnote{The data are available at the following URL: \url{https://cycling.data.tfl.gov.uk/}.}. Considering this as a network, the nodes correspond to bike sharing stations, and a directed edge between stations $i$ and $j$ is drawn if at least one ride between station $i$ and $j$ is completed within the time period considered. In this example, one week of data were considered, from 5 September until 11 September, 2018. The total number of stations used, $n=783$; the total number of undirected edges $|E|=\numprint{96060}$, implying the adjacency matrix is fairly dense. 

\begin{figure}[!b]
\centering

\begin{subfigure}[t]{0.48\textwidth}
\centering
\caption{Posterior distributions of $K_\varnothing, H_\varnothing$ for $\mvec A$}
\includegraphics[width=.97\textwidth]{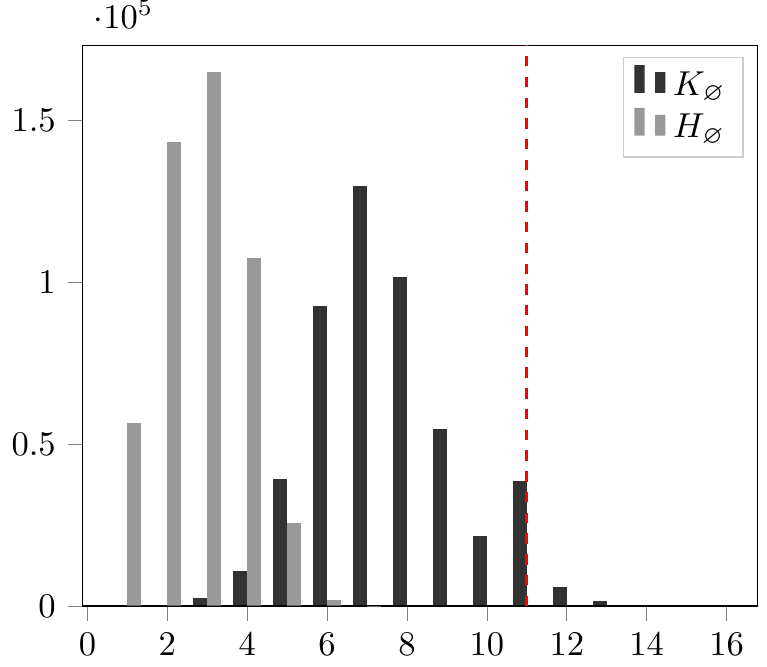}
\label{santander_hist_adj}
\end{subfigure}
\begin{subfigure}[t]{0.48\textwidth}
\centering
\caption{Posterior distributions of $K_\varnothing, H_\varnothing$ for $\mvec L$}
\includegraphics[width=.97\textwidth]{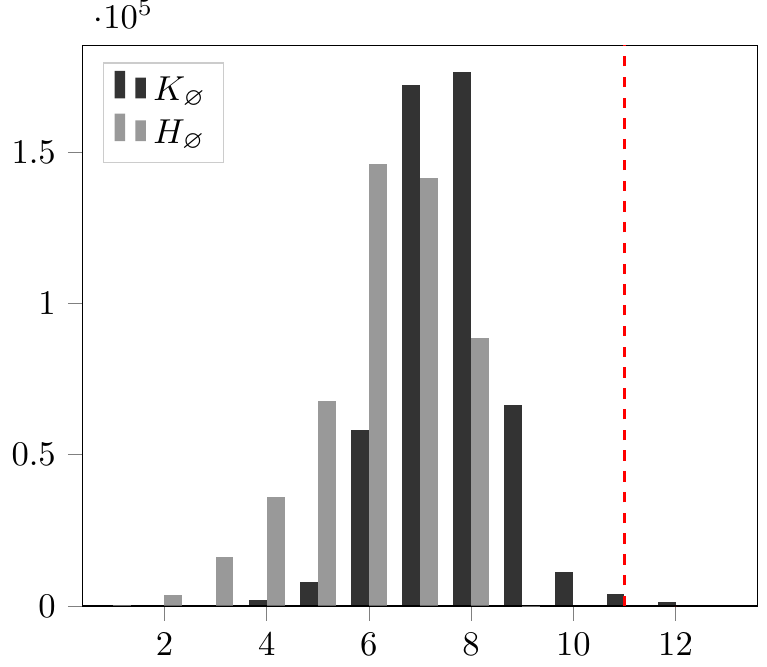}
\label{santander_hist_lap}
\end{subfigure}

\begin{subfigure}[t]{0.48\textwidth}
\centering
\caption{Magnitudes of eigenvalues of $\mvec A$}
\includegraphics[width=.97\textwidth]{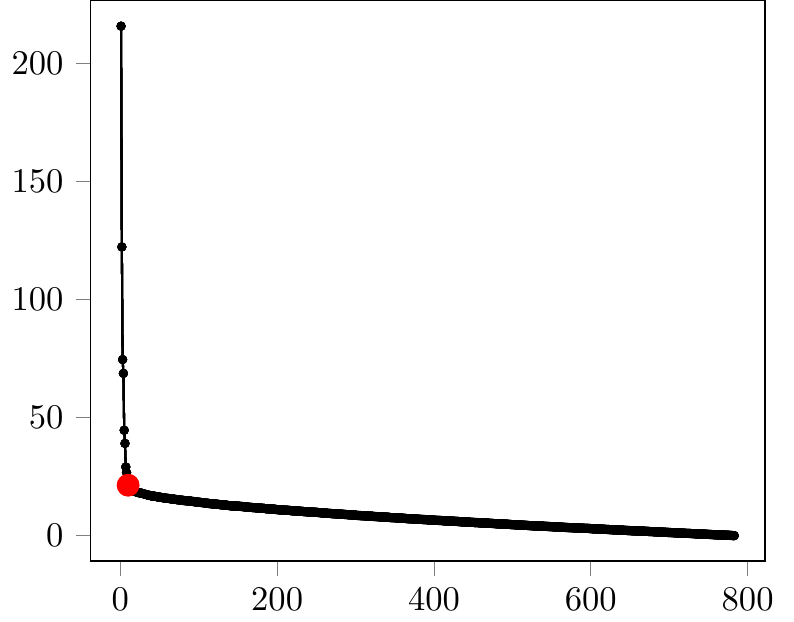}
\label{santander_eigen_adj}
\end{subfigure}
\begin{subfigure}[t]{0.48\textwidth}
\centering
\caption{Magnitudes of eigenvalues of $\mvec L$}
\includegraphics[width=.97\textwidth]{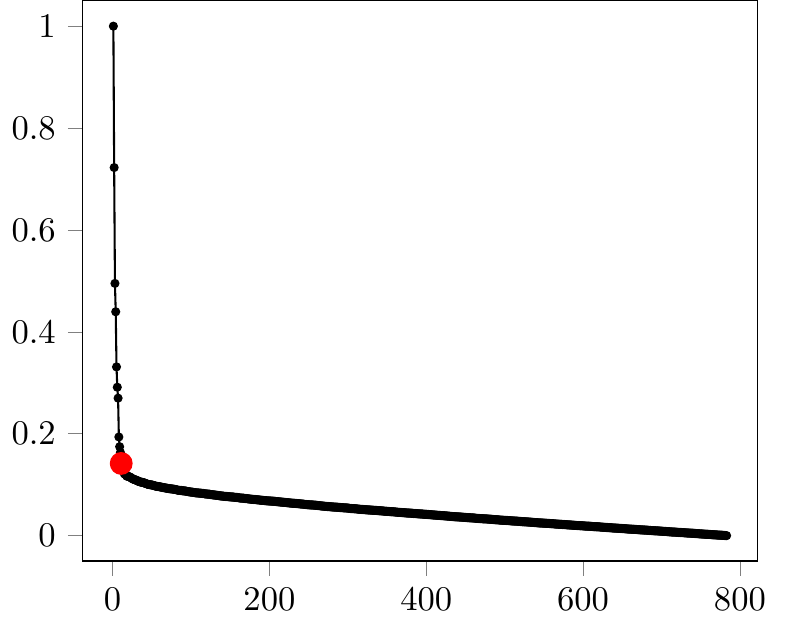}
\label{santander_eigen_lap}
\end{subfigure}

\caption{Posterior distributions and scree-plots for the Santander bike network data using adjacency and Laplacian embeddings, for the unconstrained model \eqref{eq:d_geom}. MAP estimates of $d$ are plotted in {\bf\color{red}red}.}
\label{santander_out}
\end{figure}

The results of the Bayesian inferential procedure, using the unconstrained prior \eqref{eq:d_geom} for $d$, applied to the adjacency and Laplacian embeddings for the Santander bike network are presented in Figure~\ref{santander_out}. The initial value of $K$ was set to $10$, with $m=25$, but similar estimates were obtained using different starting points for $K$ and different values of $m$. It is interesting to note the different shapes of the posterior barplots of $K_\varnothing$ and $H_\varnothing$, Figures~\ref{santander_hist_adj} and \ref{santander_hist_lap}, showing that the second-level clustering is crucial to obtain a more accurate model fit when the adjacency embedding is used. On the other hand, for the Laplacian embedding, the posteriors for $K_\varnothing$ and $H_\varnothing$ are extremely similar, suggesting that the second-level clustering is not required for $m=25$.  The \textit{maximum a posteriori} (MAP) values $d=11$ (adjacency) and $d=12$ (Laplacian) clearly correspond to the elbows in the corresponding scree-plots in Figures~\ref{santander_eigen_adj} and~\ref{santander_eigen_lap}.  

Note that, especially in the case of the adjacency embedding, $d$ and $K$ have similar values, showing that the two graphs might be well described by a stochastic blockmodel. Similarly, the constrained model with $d\overset{d}{\sim}\mathrm{Uniform}\{1,\dots,K_\varnothing\}$ \eqref{full_model} returns the same MAP estimates for $d$, but the constraint $d\leq K_\varnothing$ results in a larger number of small clusters; the posterior of $K_\varnothing$ essentially reduces to the rescaled probability mass function obtained from the unrestricted model, constrained such that $d\leq K_\varnothing$, since the posterior for $d$ is approximately a point mass. 

\begin{figure}[!t]
\centering
\includegraphics[width=\textwidth]{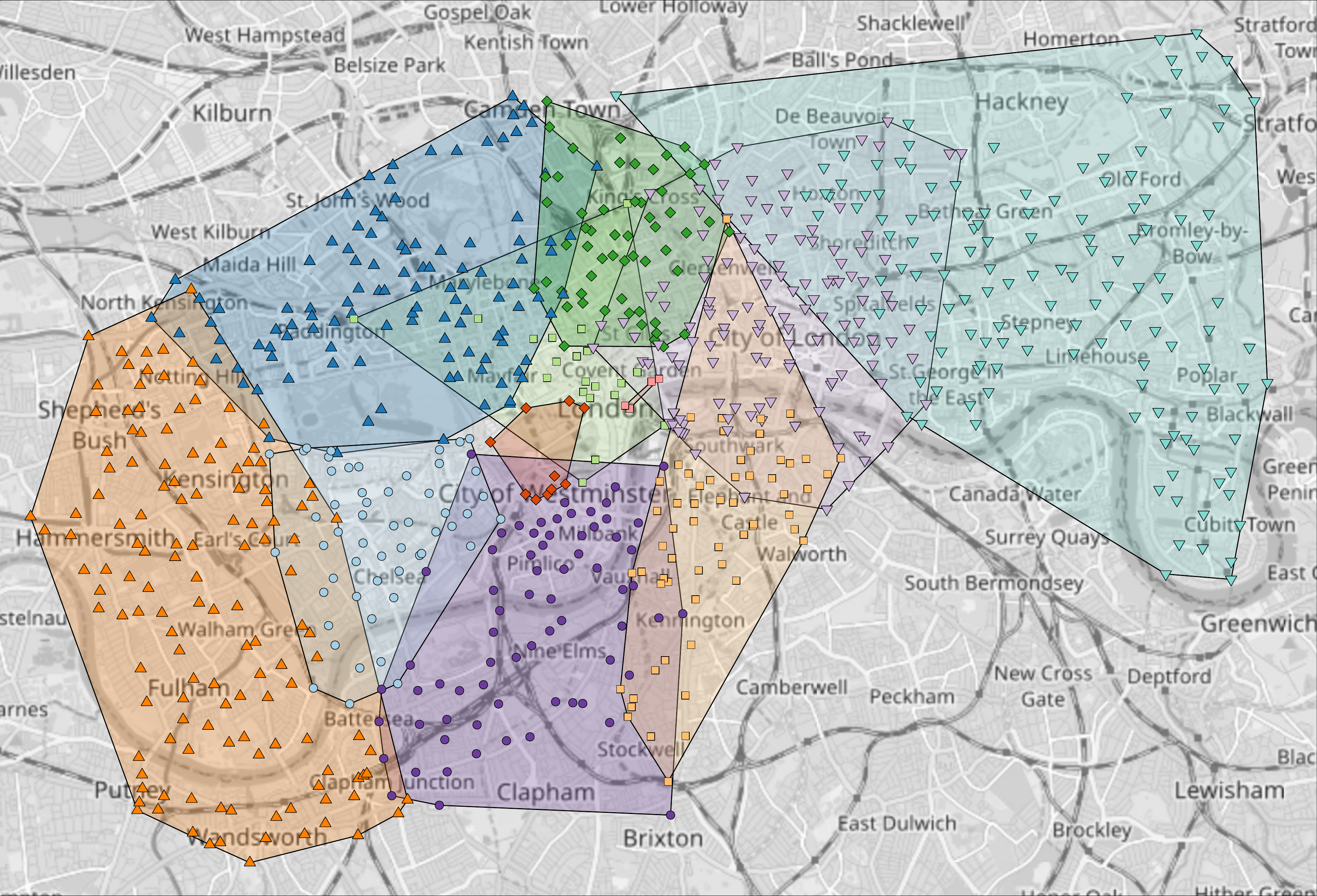}
\caption{Santander bike sharing stations in London and maximum a posteriori estimates of the cluster allocations of the stations, obtained using hierarchical clustering with distance $1-\hat\pi_{ij}$ \citep{medve}, $K=11$. Stations in the same convex hull share the same cluster.}
\label{cycle_graph}
\end{figure}

The resulting estimated clustering for the unconstrained model \eqref{eq:d_geom} based on the adjacency embedding and $K=11$ (the MAP for $d$), plotted in Figure~\ref{cycle_graph}, shows a clear structure: the largest communities have approximately the same extension, with a few exceptions. This is expected, since the bikes are free for the first $30$ minutes and have limited speed, and are therefore used for small distance journeys. Two clusters are significantly smaller than the others, and correspond to touristic areas around Westminster, Covent Garden and Buckingham Palace. On the other hand, two clusters have a large geographical extension, and cover the East and West London areas. For the adjacency embedding, the MAP clustering obtained from the restricted model is almost identical. The PEAR method \citep{fritsch} suggests $K=7$ communities instead. Similarly, if the Laplacian embedding is used, the MAP clustering structure suggested by PEAR has $K=7$ communities for the unconstrained model \eqref{eq:d_geom} for $d$ and $K=12$ for the constrained model \eqref{full_model}.

\subsection{Directed graphs: Enron Email Dataset}

Next, the algorithm is applied to a directed network: the Enron Email Dataset\footnote{The entire version of the data is available at the following URL: \url{https://www.cs.cmu.edu/~./enron/}.}. The Enron database is a corpus of emails sent by the employees of the Enron corporation. The version of the Enron data which has been analysed here is described in \cite{priebe_enron}, and consists of $n=184$ nodes and $\numprint{3010}$ directed edges. A directed edge $i\to j$ is drawn if the employee $i$ sent an email to the employee $j$. 

\begin{figure}[p]
\centering

\begin{subfigure}[t]{0.48\textwidth}
\centering
\caption{Unconstrained $d$.}
\includegraphics[width=.97\textwidth]{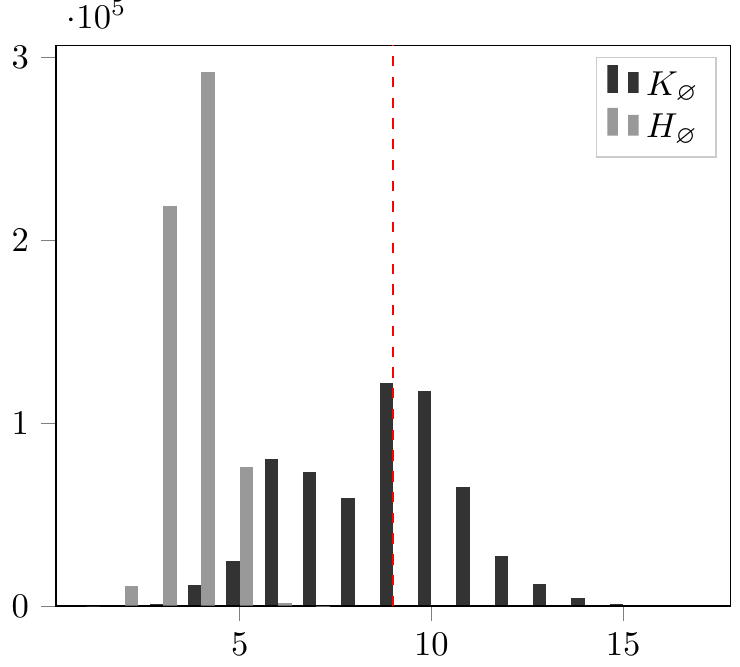}
\label{enron_hist}
\end{subfigure}
\begin{subfigure}[t]{0.48\textwidth}
\centering
\caption{Constrained $d\leq K_\varnothing$.}
\includegraphics[width=.97\textwidth]{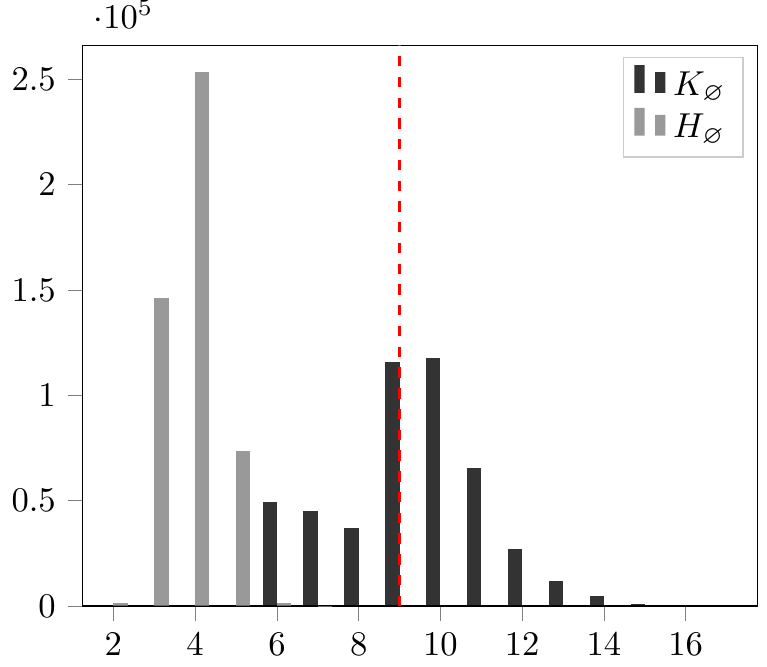}
\label{enron_hist_filter}
\end{subfigure}

\begin{subfigure}[t]{0.48\textwidth}
\centering
\caption{Unconstrained $d$, no second-level clustering.}
\includegraphics[width=.97\textwidth]{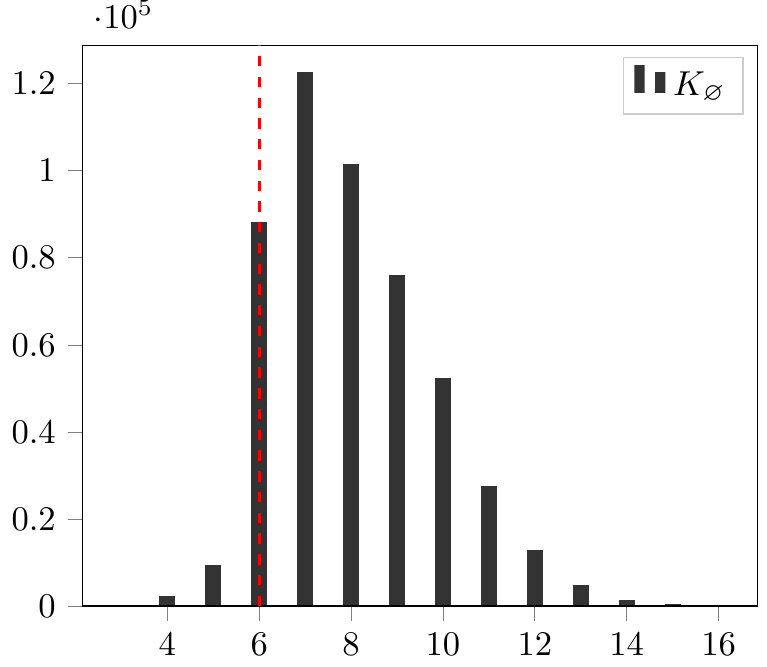}
\label{enron_hist_sord}
\end{subfigure}
\begin{subfigure}[t]{0.48\textwidth}
\centering
\caption{Constrained $d\leq K_\varnothing$, no second-level clustering.}
\includegraphics[width=.97\textwidth]{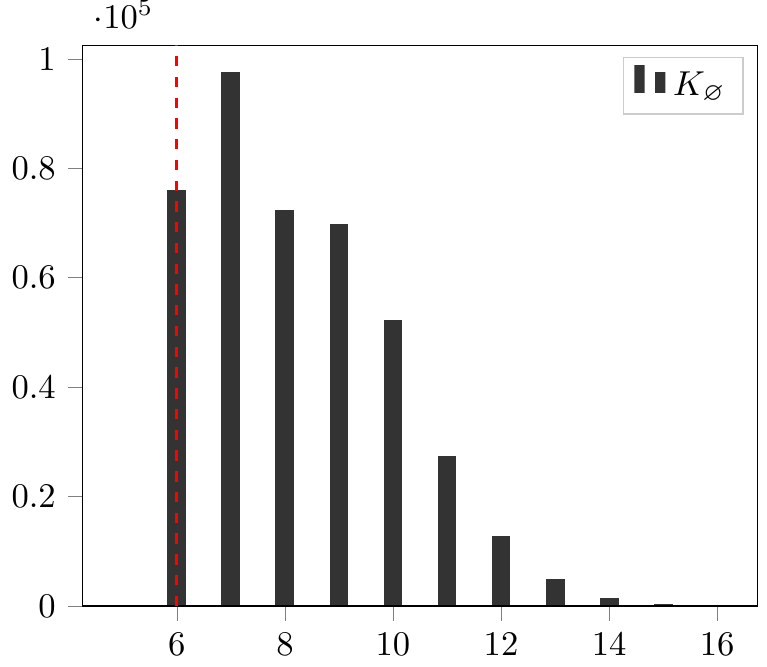}
\label{enron_hist_sord_filter}
\end{subfigure}

\centering
\begin{subfigure}[t]{0.48\textwidth}
\centering
\caption{Scree-plot of singular values of $\mvec A$}
\includegraphics[width=\textwidth]{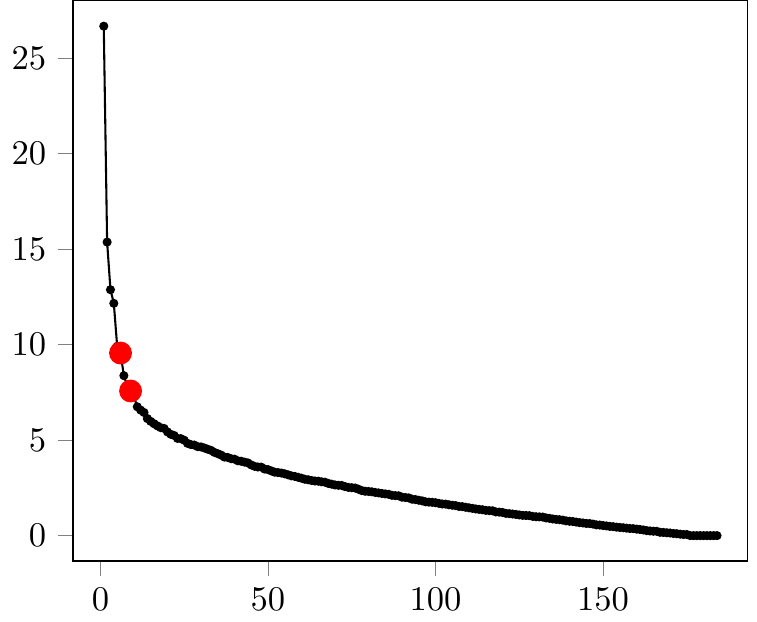}
\label{enron_eigen}
\end{subfigure}

\caption{Posterior distributions of $K_\varnothing, H_\varnothing$ for the Enron data. MAP estimates of $d$ are in {\bf\color{red} red}.} 
\label{enron_out}
\end{figure}

The results of analysing this network are presented in Figure~\ref{enron_out}. The initial value of $K$ was set to $10$, with $m=25$, but again similar results were obtained using different starting points for $K$ and different values of $m$. The plots in Figures~\ref{enron_hist} and~\ref{enron_hist_filter} report the estimated posterior distributions of $K_\varnothing$ and $H_\varnothing$ for the constrained \eqref{full_model} and unconstrained \eqref{eq:d_geom} models. Interestingly, the MAP estimate for $d$ coincides with the MAP estimate for $K$ in the unconstrained model, which is promising. For the constrained model, the MAP for $K$ exceeds the MAP for $d$ by 1, allowing for $\mathrm{rank}(\mvec B)<K$. Overall, $d$ and $K$ have similar values, showing that the graph might be well described by a directed stochastic blockmodel. 

The posterior distributions for $K_\varnothing$ and $H_\varnothing$ in Figures~\ref{enron_hist} and~\ref{enron_hist_filter} are fairly different, showing that the second-level clustering might be relevant for this model. Inference on the model without second-level clustering confirms this impression: the posteriors for $K_\varnothing$, presented in Figures~\ref{enron_hist_sord} and~\ref{enron_hist_sord_filter} have a more symmetric shape, and the MAP latent dimension is $d = 6$. As before, the MAP for $K$ is $d+1=7$, providing some evidence for the possibility $\mathrm{rank}(\mvec B)<K$.

From Figure~\ref{enron_eigen}, the selected MAP values $d=6$ and $d=9$ for the models with and without second-level clustering seem to be a tradeoff between the two most popular criteria for selection of the appropriate latent dimension: the \textit{eigengap} heuristic suggests $d=5$ if the second largest difference is considered, and the elbow in the scree-plot is approximately located around $d\approx15$. 



\section{Conclusion}

In this article, a novel Bayesian model has been proposed for automatic and simultaneous estimation of the number of communities and latent dimension of stochastic blockmodels, interpreted as special cases of generalised random dot product graphs. The Bayesian framework allows the number of communities $K$ and latent dimension $d$ to be treated as random variables, with associated posterior distributions. The postulated model is based on asymptotic results in the theory of network embeddings and random dot product graphs, and has been validated on synthetic datasets, showing good performance at recovering the latent parameters and communities. The model has been extended to directed and bipartite graphs, using SVD embeddings and allowing for co-clustering.

Overall, the main advantages of the proposed methodology is to allow for an arbitrarily large value of $m$, the number of columns (dimension) of the embedding at the first stage of the analysis, and then to treat $d$ and $K$ separately, allowing for the case $d=\mathrm{rank}(\mvec B)<K$, which is often overlooked in the literature. Problems arising from overspercification of $m$ are tackled using a second-level clustering procedure. Also, the model provides an automated procedure and criterion to select the dimension of the embedding and an appropriate number of communities. If $d$ is not constrained to be less than or equal to $K$, the model also provides empirical evidence of the goodness-of-fit of a stochastic blockmodel for the observed data. Results on real world network data sets show encouraging results in recovering the correct $d$, when compared to commonly used heuristic methods, and the community structure. 

\section*{Code and datasets}

The \textit{python} code and datasets are available at \url{https://www.github.com/fraspass/sbm}.

\section*{Acknowledgements}

The authors gratefully acknowledge funding from the EPSRC and the Heilbronn Institute for Mathematical Research.


\bibliographystyle{rss}
\singlespacing
\bibliography{biblio}

\begin{thebibliography}{87}
\expandafter\ifx\csname natexlab\endcsname\relax\def\natexlab#1{#1}\fi
\expandafter\ifx\csname url\endcsname\relax
  \def\url#1{\texttt{#1}}\fi
\expandafter\ifx\csname urlprefix\endcsname\relax\def\urlprefix{URL: }\fi

\bibitem[{Abbe(2018)}]{abbe}
Abbe, E. (2018) Community detection and stochastic block models: Recent
  developments.
\newblock \textit{Journal of Machine Learning Research}, \textbf{18}, 1--86.

\bibitem[{Abbe et~al.(2016)Abbe, Bandeira and Hall}]{abbe_recovery}
Abbe, E., Bandeira, A.~S. and Hall, G. (2016) Exact recovery in the stochastic
  block model.
\newblock \textit{IEEE Transactions on Information Theory}, \textbf{62},
  471--487.

\bibitem[{Airoldi et~al.(2008)Airoldi, Blei, Fienberg and Xing}]{airoldi}
Airoldi, E.~M., Blei, D.~M., Fienberg, S.~E. and Xing, E.~P. (2008) {Mixed
  Membership Stochastic Blockmodels}.
\newblock \textit{Journal of Machine Learning Research}, \textbf{9},
  1981--2014.

\bibitem[{Amini et~al.(2013)Amini, Chen, Bickel and Levina}]{amini}
Amini, A.~A., Chen, A., Bickel, P.~J. and Levina, E. (2013) Pseudo-likelihood
  methods for community detection in large sparse networks.
\newblock \textit{Annals of Statistics}, \textbf{41}, 2097--2122.

\bibitem[{Andrews and McNicholas(2014)}]{andrews}
Andrews, J.~L. and McNicholas, P.~D. (2014) Variable selection for clustering
  and classification.
\newblock \textit{Journal of Classification}, \textbf{31}, 136--153.

\bibitem[{Athreya et~al.(2017)Athreya, Fishkind, {Tang}, {Priebe}, Park,
  Vogelstein, Levin, Lyzinski and Qin}]{athreya}
Athreya, A., Fishkind, D.~E., {Tang}, M., {Priebe}, C.~E., Park, Y.,
  Vogelstein, J.~T., Levin, K., Lyzinski, V. and Qin, Y. (2017) Statistical
  inference on random dot product graphs: A survey.
\newblock \textit{Journal of Machine Learning Research}, \textbf{18},
  8393--8484.

\bibitem[{Bickel and Chen(2009)}]{Bickel09}
Bickel, P.~J. and Chen, A. (2009) A nonparametric view of network models and
  newman{\textendash}girvan and other modularities.
\newblock \textit{Proceedings of the National Academy of Sciences},
  \textbf{106}, 21068--21073.

\bibitem[{Bickel and Sarkar(2016)}]{Bickel16}
Bickel, P.~J. and Sarkar, P. (2016) Hypothesis testing for automated community
  detection in networks.
\newblock \textit{Journal of the Royal Statistical Society: Series B},
  \textbf{78}, 253--273.

\bibitem[{Binder(1978)}]{binder}
Binder, D.~A. (1978) {B}ayesian cluster analysis.
\newblock \textit{Biometrika}, \textbf{65}, 31--38.

\bibitem[{Bouchard-C{{\^o}}t{{\'e}} et~al.(2017)Bouchard-C{{\^o}}t{{\'e}},
  Doucet and Roth}]{particle}
Bouchard-C{{\^o}}t{{\'e}}, A., Doucet, A. and Roth, A. (2017) Particle gibbs
  split-merge sampling for {B}ayesian inference in mixture models.
\newblock \textit{Journal of Machine Learning Research}, \textbf{18}, 1--39.

\bibitem[{Cai et~al.(2018)Cai, Zheng and Chang}]{cai_survey}
Cai, H., Zheng, V.~W. and Chang, K.~C. (2018) A comprehensive survey of graph
  embedding: Problems, techniques, and applications.
\newblock \textit{IEEE Transactions on Knowledge and Data Engineering},
  \textbf{30}, 1616--1637.

\bibitem[{{Cape} et~al.(2018){Cape}, {Tang} and {Priebe}}]{Cape18}
{Cape}, J., {Tang}, M. and {Priebe}, C.~E. (2018) {On spectral embedding
  performance and elucidating network structure in stochastic block model
  graphs}.
\newblock \textit{arXiv e-prints}, arXiv:1808.04855.

\bibitem[{Celisse et~al.(2012)Celisse, Daudin and Pierre}]{celisse}
Celisse, A., Daudin, J. and Pierre, L. (2012) Consistency of maximum-likelihood
  and variational estimators in the stochastic block model.
\newblock \textit{Electronic Journal of Statistics}, \textbf{6}, 1847--1899.

\bibitem[{Chatterjee(2015)}]{chatter}
Chatterjee, S. (2015) Matrix estimation by universal singular value
  thresholding.
\newblock \textit{Annals of Statistics}, \textbf{43}, 177--214.

\bibitem[{Chen and Lei(2018)}]{Chen18}
Chen, K. and Lei, J. (2018) Network cross-validation for determining the number
  of communities in network data.
\newblock \textit{Journal of the American Statistical Association},
  \textbf{113}, 241--251.

\bibitem[{Dahl(2003)}]{dahl}
Dahl, D.~B. (2003) An improved merge-split sampler for conjugate {D}irichlet
  process mixture models.
\newblock \textit{Tech. Rep. 1086}, Department of Statistics, University of
  Wisconsin, Madison.

\bibitem[{Dahl(2006)}]{dahl_clust}
--- (2006) \textit{Model-Based Clustering for Expression Data via a {D}irichlet
  Process Mixture Model}, 201--218.
\newblock Cambridge University Press.

\bibitem[{Dellaportas and Papageorgiou(2006)}]{dellaportas}
Dellaportas, P. and Papageorgiou, I. (2006) Multivariate mixtures of normals
  with unknown number of components.
\newblock \textit{Statistics and Computing}, \textbf{16}, 57--68.

\bibitem[{Dhillon(2001)}]{dhillon}
Dhillon, I.~S. (2001) Co-clustering documents and words using bipartite
  spectral graph partitioning.
\newblock In \textit{Proceedings of the Seventh ACM SIGKDD International
  Conference on Knowledge Discovery and Data Mining}, KDD '01, 269--274. New
  {Y}ork, NY, USA: ACM.

\bibitem[{Fishkind et~al.(2013)Fishkind, Sussman, Tang, Vogelstein and
  Priebe}]{Fishkind13}
Fishkind, D.~E., Sussman, D.~L., Tang, M., Vogelstein, J.~T. and Priebe, C.~E.
  (2013) Consistent adjacency-spectral partitioning for the stochastic block
  model when the model parameters are unknown.
\newblock \textit{SIAM Journal on Matrix Analysis and Applications},
  \textbf{34}, 23--39.

\bibitem[{Fowlkes et~al.(1988)Fowlkes, Gnanadesikan and Kettenring}]{fowlkes}
Fowlkes, E.~B., Gnanadesikan, R. and Kettenring, J.~R. (1988) Variable
  selection in clustering.
\newblock \textit{Journal of Classification}, \textbf{5}, 205--228.

\bibitem[{Fraley and Raftery(2002)}]{raftery}
Fraley, C. and Raftery, A.~E. (2002) Model-based clustering, discriminant
  analysis, and density estimation.
\newblock \textit{Journal of the American Statistical Association},
  \textbf{97}, 611--631.

\bibitem[{Franco~Salda{\~{n}}a et~al.(2017)Franco~Salda{\~{n}}a, Yu and
  Feng}]{Saldana17}
Franco~Salda{\~{n}}a, D., Yu, Y. and Feng, Y. (2017) How many communities are
  there?
\newblock \textit{Journal of Computational and Graphical Statistics},
  \textbf{26}, 171--181.

\bibitem[{Fritsch and Ickstadt(2009)}]{fritsch}
Fritsch, A. and Ickstadt, K. (2009) Improved criteria for clustering based on
  the posterior similarity matrix.
\newblock \textit{{B}ayesian Analysis}, \textbf{4}, 367--391.

\bibitem[{Green(1995)}]{Green95}
Green, P.~J. (1995) Reversible jump {M}arkov {C}hain {M}onte {C}arlo
  computation and {B}ayesian model determination.
\newblock \textit{Biometrika}, \textbf{82}, 711--732.

\bibitem[{Handcock et~al.(2007)Handcock, Raftery and Tantrum}]{handcock}
Handcock, M.~S., Raftery, A.~E. and Tantrum, J.~M. (2007) Model-based
  clustering for social networks.
\newblock \textit{Journal of the Royal Statistical Society: Series A
  (Statistics in Society)}, \textbf{170}, 301--354.

\bibitem[{Hoff et~al.(2002)Hoff, Raftery and Handcock}]{hoff}
Hoff, P.~D., Raftery, A.~E. and Handcock, M.~S. (2002) Latent space approaches
  to social network analysis.
\newblock \textit{Journal of the American Statistical Association},
  \textbf{97}, 1090--1098.

\bibitem[{Holland et~al.(1983)Holland, Laskey and Leinhardt}]{holland}
Holland, P.~W., Laskey, K.~B. and Leinhardt, S. (1983) Stochastic blockmodels:
  First steps.
\newblock \textit{Social Networks}, \textbf{5}, 109 -- 137.

\bibitem[{Jain and Neal(2004)}]{jain}
Jain, S. and Neal, R.~M. (2004) A split-merge {M}arkov {C}hain {M}onte {C}arlo
  procedure for the {D}irichlet process mixture model.
\newblock \textit{Journal of Computational and Graphical Statistics},
  \textbf{13}, 158--182.

\bibitem[{Jasra et~al.(2005)Jasra, Holmes and Stephens}]{jasra}
Jasra, A., Holmes, C.~C. and Stephens, D.~A. (2005) {M}arkov {C}hain {M}onte
  {C}arlo methods and the label switching problem in {B}ayesian mixture
  modeling.
\newblock \textit{Statistical Science}, \textbf{20}, 50--67.

\bibitem[{Jolliffe(2002)}]{jolliffe}
Jolliffe, I.~T. (2002) \textit{Principal Component Analysis}.
\newblock Springer Series in Statistics. Springer.

\bibitem[{Karrer and Newman(2011)}]{karrer}
Karrer, B. and Newman, M. E.~J. (2011) Stochastic blockmodels and community
  structure in networks.
\newblock \textit{Physical Review E}, \textbf{83}.

\bibitem[{Krivitsky et~al.(2009)Krivitsky, Handcock, Raftery and
  Hoff}]{krivitsky}
Krivitsky, P.~N., Handcock, M.~S., Raftery, A.~E. and Hoff, P.~D. (2009)
  Representing degree distributions, clustering, and homophily in social
  networks with latent cluster random effects models.
\newblock \textit{Social Networks}, \textbf{31}, 204--213.

\bibitem[{Lau and Green(2007)}]{lau}
Lau, J.~W. and Green, P.~J. (2007) {B}ayesian model-based clustering
  procedures.
\newblock \textit{Journal of Computational and Graphical Statistics},
  \textbf{16}, 526--558.

\bibitem[{Law et~al.(2004)Law, Figueiredo and Jain}]{law}
Law, M. H.~C., Figueiredo, M. A.~T. and Jain, A.~K. (2004) Simultaneous feature
  selection and clustering using mixture models.
\newblock \textit{IEEE Transactions on Pattern Analysis and Machine
  Intelligence}, \textbf{26}, 1154--1166.

\bibitem[{Lei(2016)}]{Lei16}
Lei, J. (2016) A goodness-of-fit test for stochastic block models.
\newblock \textit{The Annals of Statistics}, \textbf{44}, 401--424.

\bibitem[{Lei and Rinaldo(2015)}]{lei}
Lei, J. and Rinaldo, A. (2015) Consistency of spectral clustering in stochastic
  block models.
\newblock \textit{Annals of Statistics}, \textbf{43}, 215--237.

\bibitem[{Liu(1994)}]{liu}
Liu, J.~S. (1994) The collapsed {G}ibbs sampler in {B}ayesian computations with
  applications to a gene regulation problem.
\newblock \textit{Journal of the American Statistical Association},
  \textbf{89}, 958--966.

\bibitem[{Ludkin et~al.(2018)Ludkin, Eckley and Neal}]{ludkin}
Ludkin, M., Eckley, I. and Neal, P. (2018) Dynamic stochastic block models:
  parameter estimation and detection of changes in community structure.
\newblock \textit{Statistics and Computing}, \textbf{28}, 1201--1213.

\bibitem[{Lyzinski et~al.(2014)Lyzinski, Sussman, Tang, Athreya and
  Priebe}]{lyzinski}
Lyzinski, V., Sussman, D.~L., Tang, M., Athreya, A. and Priebe, C.~E. (2014)
  Perfect clustering for stochastic blockmodel graphs via adjacency spectral
  embedding.
\newblock \textit{Electronic Journal of Statistics}, \textbf{8}, 2905--2922.

\bibitem[{Lyzinski et~al.(2017)Lyzinski, Tang, Athreya, Park and
  Priebe}]{lyz_community}
Lyzinski, V., Tang, M., Athreya, A., Park, Y. and Priebe, C.~E. (2017)
  Community detection and classification in hierarchical stochastic
  blockmodels.
\newblock \textit{IEEE Transactions on Network Science and Engineering},
  \textbf{4}, 13--26.

\bibitem[{Malliaros and Vazirgiannis(2013)}]{malliaros}
Malliaros, F.~D. and Vazirgiannis, M. (2013) Clustering and community detection
  in directed networks: A survey.
\newblock \textit{Physics Reports}, \textbf{533}, 95 -- 142.

\bibitem[{Matias and Miele(2017)}]{matias}
Matias, C. and Miele, V. (2017) Statistical clustering of temporal networks
  through a dynamic stochastic block model.
\newblock \textit{Journal of the Royal Statistical Society: Series B},
  \textbf{79}, 1119--1141.

\bibitem[{Maugis et~al.(2009)Maugis, Celeux and Martin-Magniette}]{maugis}
Maugis, C., Celeux, G. and Martin-Magniette, M.~L. (2009) {V}ariable selection
  for clustering with {G}aussian mixture models.
\newblock \textit{Biometrics}, \textbf{65}, 701--709.

\bibitem[{Medvedovic et~al.(2004)Medvedovic, Yeung and Bumgarner}]{medve}
Medvedovic, M., Yeung, K.~Y. and Bumgarner, R.~E. (2004) {B}ayesian mixture
  model based clustering of replicated microarray data.
\newblock \textit{Bioinformatics}, \textbf{20}, 1222--1232.

\bibitem[{Mengersen and Robert(1996)}]{mengersen}
Mengersen, K. and Robert, C. (1996) Testing for mixtures: a {B}ayesian entropic
  approach (with discussion).
\newblock In \textit{{B}ayesian Statistics} (eds. J.~Berger, J.~Bernardo,
  A.~Dawid, D.~Lindley and A.~Smith). Oxford University Press.

\bibitem[{Miller and Harrison(2014)}]{harrison}
Miller, J.~W. and Harrison, M.~T. (2014) Inconsistency of {P}itman-{Y}or
  process mixtures for the number of components.
\newblock \textit{Journal of Machine Learning Research}, \textbf{15},
  3333--3370.

\bibitem[{Miller and Harrison(2018)}]{miller_mix}
--- (2018) Mixture models with a prior on the number of components.
\newblock \textit{Journal of the American Statistical Association},
  \textbf{113}, 340--356.

\bibitem[{Murphy(2007)}]{murphy}
Murphy, K.~P. (2007) Conjugate {B}ayesian analysis of the gaussian
  distribution.
\newblock \textit{Tech. rep.}

\bibitem[{Newman and Reinert(2016)}]{reinert}
Newman, M. E.~J. and Reinert, G. (2016) Estimating the number of communities in
  a network.
\newblock \textit{Physical Review Letters}, \textbf{117}.

\bibitem[{Nickel(2006)}]{Nickel06}
Nickel, C. L.~M. (2006) \textit{Random dot product graphs: a model for social
  networks}.
\newblock Ph.D. thesis, The Johns Hopkins University.

\bibitem[{Nobile(2004)}]{nobile}
Nobile, A. (2004) On the posterior distribution of the number of components in
  a finite mixture.
\newblock \textit{Annals of Statistics}, \textbf{32}, 2044--2073.

\bibitem[{Nobile and Fearnside(2007)}]{nobile_fearn}
Nobile, A. and Fearnside, A.~T. (2007) {B}ayesian finite mixtures with an
  unknown number of components: The allocation sampler.
\newblock \textit{Statistics and Computing}, \textbf{17}, 147--162.

\bibitem[{Nowicki and Snijders(2001)}]{nowi}
Nowicki, K. and Snijders, T. A.~B. (2001) Estimation and prediction for
  stochastic blockstructures.
\newblock \textit{Journal of the American Statistical Association},
  \textbf{96}, 1077--1087.

\bibitem[{Peixoto(2018)}]{peixoto}
Peixoto, T.~P. (2018) {B}ayesian stochastic blockmodeling.
\newblock In \textit{Advances in Network Clustering and Blockmodeling} (eds.
  P.~Doreian, V.~Batagelj and A.~Ferligoj). New {Y}ork, NY, USA: Wiley.

\bibitem[{{Pensky} and {Zhang}(2017)}]{pensky}
{Pensky}, M. and {Zhang}, T. (2017) {Spectral clustering in the dynamic
  stochastic block model}.
\newblock \textit{arXiv e-prints}.

\bibitem[{Priebe et~al.(2005)Priebe, Conroy, Marchette and Park}]{priebe_enron}
Priebe, C.~E., Conroy, J.~M., Marchette, D.~J. and Park, Y. (2005) Scan
  statistics on enron graphs.
\newblock \textit{Computational {\&} Mathematical Organization Theory},
  \textbf{11}, 229--247.

\bibitem[{Priebe et~al.(2019)Priebe, Park, Vogelstein, Conroy, Lyzinski, Tang,
  Athreya, Cape and Bridgeford}]{Priebe19}
Priebe, C.~E., Park, Y., Vogelstein, J.~T., Conroy, J.~M., Lyzinski, V., Tang,
  M., Athreya, A., Cape, J. and Bridgeford, E. (2019) On a two-truths
  phenomenon in spectral graph clustering.
\newblock \textit{Proceedings of the National Academy of Sciences},
  \textbf{116}, 5995--6000.

\bibitem[{Raftery and Dean(2006)}]{dean}
Raftery, A.~E. and Dean, N. (2006) Variable selection for model-based
  clustering.
\newblock \textit{Journal of the American Statistical Association},
  \textbf{101}, 168--178.

\bibitem[{Rand(1971)}]{rand}
Rand, W.~M. (1971) Objective criteria for the evaluation of clustering methods.
\newblock \textit{Journal of the American Statistical Association},
  \textbf{66}, 846--850.

\bibitem[{Rastelli et~al.(2018)Rastelli, Latouche and Friel}]{rastelli}
Rastelli, R., Latouche, P. and Friel, N. (2018) Choosing the number of groups
  in a latent stochastic blockmodel for dynamic networks.
\newblock \textit{Network Science}, 1--25.

\bibitem[{Richardson and Green(1997)}]{richardson}
Richardson, S. and Green, P.~J. (1997) On {B}ayesian analysis of mixtures with
  an unknown number of components (with discussion).
\newblock \textit{Journal of the Royal Statistical Society: Series B},
  \textbf{59}, 731--792.

\bibitem[{Riolo et~al.(2017)Riolo, Cantwell, Reinert and Newman}]{riolo}
Riolo, M.~A., Cantwell, G.~T., Reinert, G. and Newman, M. E.~J. (2017)
  Efficient method for estimating the number of communities in a network.
\newblock \textit{Physical Review E}, \textbf{96}.

\bibitem[{Rohe et~al.(2011)Rohe, Chatterjee and Yu}]{rohe}
Rohe, K., Chatterjee, S. and Yu, B. (2011) Spectral clustering and the
  high-dimensional stochastic blockmodel.
\newblock \textit{Annals of Statistics}, \textbf{39}, 1878--1915.

\bibitem[{Rohe et~al.(2016)Rohe, Qin and Yu}]{scbm}
Rohe, K., Qin, T. and Yu, B. (2016) Co-clustering directed graphs to discover
  asymmetries and directional communities.
\newblock \textit{Proceedings of the National Academy of Sciences}.

\bibitem[{Rubin-Delanchy et~al.(2016)Rubin-Delanchy, Adams and
  Heard}]{disassortativity}
Rubin-Delanchy, P., Adams, N.~M. and Heard, N.~A. (2016) Disassortativity of
  computer networks.
\newblock In \textit{2016 IEEE Conference on Intelligence and Security
  Informatics (ISI)}, 243--247.

\bibitem[{{Rubin-Delanchy} et~al.(2017){Rubin-Delanchy}, {Priebe}, {Tang} and
  {Cape}}]{prd}
{Rubin-Delanchy}, P., {Priebe}, C.~E., {Tang}, M. and {Cape}, J. (2017) {A
  statistical interpretation of spectral embedding: the generalised random dot
  product graph}.
\newblock \textit{ArXiv e-prints}.

\bibitem[{Sengupta and Chen(2018)}]{sengupta}
Sengupta, S. and Chen, Y. (2018) A block model for node popularity in networks
  with community structure.
\newblock \textit{Journal of the Royal Statistical Society: Series B},
  \textbf{80}, 365--386.

\bibitem[{Snijders and Nowicki(1997)}]{snijders}
Snijders, T. A.~B. and Nowicki, K. (1997) Estimation and prediction for
  stochastic blockmodels for graphs with latent block structure.
\newblock \textit{Journal of Classification}, \textbf{14}, 75--100.

\bibitem[{Stephens(2000)}]{stephens}
Stephens, M. (2000) {B}ayesian analysis of mixture models with an unknown
  number of components--an alternative to reversible jump methods.
\newblock \textit{The Annals of Statistics}, \textbf{28}, 40--74.

\bibitem[{Sussman et~al.(2012)Sussman, Minh, Fishkind and Priebe}]{sussman}
Sussman, D.~L., Minh, T., Fishkind, D.~E. and Priebe, C.~E. (2012) A consistent
  adjacency spectral embedding for stochastic blockmodel graphs.
\newblock \textit{Journal of the American Statistical Association},
  \textbf{107}, 1119--1128.

\bibitem[{Sussman et~al.(2014)Sussman, Tang and Priebe}]{sussman_ieee}
Sussman, D.~L., Tang, M. and Priebe, C.~E. (2014) Consistent latent position
  estimation and vertex classification for random dot product graphs.
\newblock \textit{IEEE Transactions on Pattern Analysis and Machine
  Intelligence}, \textbf{36}, 48--57.

\bibitem[{Tang and Priebe(2018)}]{tang}
Tang, M. and Priebe, C.~E. (2018) Limit theorems for eigenvectors of the
  normalized laplacian for random graphs.
\newblock \textit{Annals of Statistics}, \textbf{46}, 2360--2415.

\bibitem[{Tang et~al.(2013)Tang, Sussman and Priebe}]{Tang13}
Tang, M., Sussman, D.~L. and Priebe, C.~E. (2013) Universally consistent vertex
  classification for latent positions graphs.
\newblock \textit{Annals of Statistics}, \textbf{41}, 1406--1430.

\bibitem[{{van der Pas} and {van der Vaart}(2018)}]{vdv}
{van der Pas}, S.~L. and {van der Vaart}, A.~W. (2018) {B}ayesian community
  detection.
\newblock \textit{{B}ayesian Analysis}, \textbf{13}, 767--796.

\bibitem[{{von Luxburg}(2007)}]{spectral}
{von Luxburg}, U. (2007) A tutorial on spectral clustering.
\newblock \textit{Statistics and Computing}, \textbf{1}, 395--416.

\bibitem[{Wang and Wong(1987)}]{wang_wong}
Wang, W.~J. and Wong, G.~Y. (1987) Stochastic blockmodels for directed graphs.
\newblock \textit{Journal of the American Statistical Association},
  \textbf{82}, 8--19.

\bibitem[{Wang and Bickel(2017)}]{wang_model_selection}
Wang, Y. X.~R. and Bickel, P.~J. (2017) Likelihood-based model selection for
  stochastic block models.
\newblock \textit{Annals of Statistics}, \textbf{45}, 500--528.

\bibitem[{{Wolfe} and {Olhede}(2013)}]{olhede}
{Wolfe}, P.~J. and {Olhede}, S.~C. (2013) Nonparametric graphon estimation.
\newblock \textit{arXiv e-prints}.

\bibitem[{Xu and {Hero III}(2013)}]{xu}
Xu, K.~S. and {Hero III}, A.~O. (2013) Dynamic stochastic blockmodels:
  Statistical models for time-evolving networks.
\newblock In \textit{Social Computing, Behavioral-Cultural Modeling and
  Prediction} (eds. A.~Greenberg, W.~Kennedy and N.~Bos), 201--210. Berlin,
  Heidelberg: Springer Berlin Heidelberg.

\bibitem[{{Yang} et~al.(2019){Yang}, {Priebe}, {Park} and {Marchette}}]{Yang19}
{Yang}, C., {Priebe}, C.~E., {Park}, Y. and {Marchette}, D.~J. (2019)
  Simultaneous dimensionality and complexity model selection for spectral graph
  clustering.
\newblock \textit{arXiv e-prints}, arXiv:1904.02926.

\bibitem[{Young and Scheinerman(2007)}]{young_rdpg}
Young, S.~J. and Scheinerman, E.~R. (2007) Random dot product graph models for
  social networks.
\newblock In \textit{Algorithms and Models for the Web-Graph} (eds. A.~Bonato
  and F.~R.~K. Chung), 138--149. Berlin, Heidelberg: Springer Berlin
  Heidelberg.

\bibitem[{Young and Scheinerman(2008)}]{young}
--- (2008) Directed random dot product graphs.
\newblock \textit{Internet Mathematics}, \textbf{5}, 91--112.

\bibitem[{Zhang et~al.(2004)Zhang, Chan, Wu and Chen}]{zhang}
Zhang, Z., Chan, K.~L., Wu, Y. and Chen, C. (2004) Learning a multivariate
  gaussian mixture model with the reversible jump mcmc algorithm.
\newblock \textit{Statistics and Computing}, \textbf{14}, 343--355.

\bibitem[{Zhao et~al.(2011)Zhao, Levina and Zhu}]{Zhao11}
Zhao, Y., Levina, E. and Zhu, J. (2011) Community extraction for social
  networks.
\newblock \textit{Proceedings of the National Academy of Sciences},
  \textbf{108}, 7321--7326.

\bibitem[{Zheng and Skillicorn(2015)}]{zheng}
Zheng, Q. and Skillicorn, D.~B. (2015) Spectral embedding of directed networks.
\newblock \textit{2015 IEEE/ACM International Conference on Advances in Social
  Networks Analysis and Mining (ASONAM)}, 432--439.

\bibitem[{Zhu and Ghodsi(2006)}]{zhu}
Zhu, M. and Ghodsi, A. (2006) Automatic dimensionality selection from the scree
  plot via the use of profile likelihood.
\newblock \textit{Computational Statistics \& Data Analysis}, \textbf{51}, 918
  -- 930.

\end{thebibliography}

\end{document}